\documentclass{article}

\usepackage{arxiv}

\usepackage[utf8]{inputenc} 
\usepackage[T1]{fontenc}    
\usepackage[pdfencoding=auto, psdextra]{hyperref}

\usepackage{url}            
\usepackage{booktabs}       
\usepackage{amsfonts}       
\usepackage{nicefrac}       
\usepackage{microtype}      
\usepackage{lipsum}		
\usepackage{graphicx}
\usepackage[numbers,sort&compress]{natbib}

\usepackage{doi}

\usepackage{float}

\usepackage{amsmath,amssymb,amsfonts}
\usepackage{relsize}

\usepackage{yhmath}

\usepackage[capitalise]{cleveref}

\usepackage{graphicx}
\graphicspath{{Figures-pdf/}}
\usepackage{todonotes}

\newcommand*{\tran}{^{\mkern-1.5mu\mathsf{T}}}
\newcommand*{\mfon}{^{^{^{\hspace{-7pt}\text{\scriptsize MFON}}}}\hspace{-14pt}}
\newcommand*{\gpod}{^{^{^{\hspace{-7pt}\text{\scriptsize GPOD}}}}\hspace{-13pt}}
\newcommand*{\ar}{^{^{^{\hspace{-7pt}\text{\scriptsize Autoregressive}}}}\hspace{-31pt}}
\newcommand*{\inc}{^{^{^{\hspace{-7pt}\text{\scriptsize Incremental}}}}\hspace{-23pt}}

\title{A Multifidelity Deep Operator Network Approach to Closure for Multiscale Systems}

\author{Shady E.~Ahmed \\ 
	Advanced Computing, Mathematics and Data Division \\ 
 Pacific Northwest National Laboratory\\
	Richland, WA 99354 \\
	\texttt{shady.ahmed@pnnl.gov} \\
	\And
	Panos Stinis \\
              Advanced Computing, Mathematics and Data Division \\Pacific Northwest National Laboratory\\
	Richland, WA 99354 \\
}

\renewcommand{\shorttitle}{Multifidelity Operator Learning for Multiscale Systems Closure}

\hypersetup{
pdftitle={A Multifidelity Deep Operator Network Approach to Closure for Multiscale Systems},
pdfauthor={Shady E.~Ahmed, Panos Stinis},
pdfkeywords= {Reduced order models, DeepONet, In-the-loop training, Differentiable physics, Multifidelity learning},
}

\begin{document}
\maketitle

\begin{abstract}
Projection-based reduced order models (PROMs) have shown promise  in representing the behavior of multiscale systems using a small set of generalized (or latent) variables. Despite their success, PROMs can be susceptible to inaccuracies, even instabilities, due to the improper accounting of the interaction between the resolved and unresolved scales of the multiscale system (known as the closure problem). In the current work, we interpret closure as a multifidelity problem and use a multifidelity deep operator network (DeepONet) framework to address it. In addition, to enhance the stability and accuracy of the  multifidelity-based closure, we employ the recently developed ``in-the-loop" training approach from the literature on coupling physics and machine learning models. The resulting approach is tested on shock advection for the one-dimensional viscous Burgers equation and vortex merging using the two-dimensional Navier-Stokes equations. The numerical experiments show  significant improvement of the predictive ability of the closure-corrected PROM  over the un-corrected one both in the interpolative and the extrapolative regimes.
\end{abstract}


\section{Introduction} \label{sec:intro}
High fidelity simulations produce invaluable information to augment our understanding of the world and physical processes around us. However, their use has been limited in multi-query and outer-loop applications such as design optimization, model predictive control, and uncertainty quantification. This is simply because they are too computationally demanding and we do not often have the computing infrastructure that enables multiple forward runs within the allowable turnaround time. Therefore, there is a need to build computationally-lightweight models that describe the system's behavior with acceptable accuracy. The last few decades have witnessed increased interest in model order reduction (MOR) developments. Among these, projection-based reduced order models (PROMs) have shown promise  in representing the behavior of multiscale systems using a small set of generalized (or latent) variables. The derivation of PROM commonly involves two steps: (1) constructing a few basis functions that encapsulate the dominant features of the system, and (2) defining a model to estimate the leading coefficients (weights) of these basis functions at different times/parameters as they are used to expand the solution of the system under investigation.

The combination of proper orthogonal decomposition (POD) \cite{holmes2012turbulence} and Galerkin methods \cite{noack2011reduced} has been a main driver for PROM developments in fluid dynamics, structural mechanics, and other fields. The basic idea of POD is to represent a data set as a linear combination of its dominant modes, which are calculated from the data set itself. These modes are hierarchically sorted, based on the eigenvalues of an appropriate covariance operator, signifying the relative importance of individual modes to the high dimensional data reconstruction in the $\ell_2$-sense. An underlying assumption of scale-separation is often implied by considering the analogy to Fourier basis wherein the low-index modes represent the largest scales while the high-index modes correspond to the smaller scales \cite{couplet2003intermodal}. To reduce the computational complexity of the system, only a handful of the leading POD modes are retained while the remaining ones are truncated. In the second step of Galerkin method, the high fidelity governing equations are projected onto the span of the selected POD modes to derive a reduced set of dynamical equations that defines the Galerkin-POD (GPOD) model. 

Although GPOD models perform well in many quasi-periodic and statistically steady state cases, they usually fail in long-time predictions especially for systems whose solution is convection-dominated as we all as for systems representing turbulent flows. {One source of the failure modes in GPOD predictions is related to their lack of stability guarantees and this line of research has been at the center of several research efforts. \citet{barone2009stable} demonstrated that the stability of PROM predictions is closely tied to the type of inner product used to define the projection. In particular, carefully-designed inner products that preserve symmetry and satisfy boundary conditions have been shown to provide better performance in terms of stability, compared to standard $L^2$ inner product. Similar findings have been reported in \cite{aubry1992spatio,iollo2000stability,xie2020lagrangian} using other forms of inner products.}

The inaccuracy (and even instability) of the GPOD models can also be attributed to the severe truncation of the POD modes. Although the small scales themselves might not be of interest, the error due to the neglect of these scales can grow rapidly and infect the accuracy of the prediction at the larger scales. Lorenz \cite{lorenz1969predictability} attributed the finite-time weather predictability barrier to the propagation of errors at small scales to large scales, coined as the real butterfly effect \cite{palmer2014real}. 
{Subspace rotation techniques (e.g., using a mixture of large energetic and small dissipative modes \cite{balajewicz2012stabilization,balajewicz2013low,balajewicz2016minimal}) have been shown to effectively address the modal truncation effects.} Compensating accurately for the contribution of the truncated scales onto the resolved ones (as a function of the resolved state variables and parameters) is known as the ``closure problem'' \cite{ahmed2021closures}. A large body of literature has focused on developing closure models in multiscale phenomena, using physical arguments \cite{aubry1988dynamics,sirisup2004spectral,borggaard2011artificial} and mathematical formulations \cite{chorin2007problem,xie2017approximate,gunzburger2017ensemble}. More recently, there has been a surge in adopting machine learning (ML) tools to build novel closure models, e.g., \cite{san2018neural,pan2018data,pawar2019deep,ahmed2021closures,gupta2021neural,bruna2022neural,linot2023stabilized}. However, most of these developments have two fundamental issues that limit their applicability (and we aim to address in this study) as follows:

\begin{enumerate}
    \item They mostly rely on deep neural network (DNN) capabilities as universal approximators of arbitrary functions. Nonetheless, a common theme in DNN-based implementations is their limited applicability when it comes to variations in physical parameters as well as initial and boundary conditions. There have been some successes in adopting DNNs for time-dependent parametric PROMs, e.g., using an ensemble of DNNs trained on clustered regions of interest in the parameter space {or feeding the neural network with extra information about the system. For instance, \citet{xie2020closure} trained a residual neural network (ResNet) to learn the closure terms in GPOD models and enriched the input vector with parameter values to improve the predictive capabilities in parametric settings.} However, the critical view of the need to re-train DNNs for new cases, specifically in extrapolative contexts, has been lingering unaddressed rigorously in the scientific community for a while. 
    
    \item Previous works on ML-based predictions often employ an idealized training environment that does not reflect the actual operation. For instance, the DNN models are usually trained as standalone components using \emph{clean and curated} data (e.g., noise-free) as opposed to the testing/deployment conditions where inputs are inevitably contaminated with all sorts of errors, e.g., measurement noise and numerical approximations. Furthermore, the ML prediction at one time step has no effect on the input data at the next time steps during the training phase. However, during testing, the inaccuracies at one step manifest themselves in the following steps. This can lead to severe inaccuracy, and even instability, when the ML model is coupled with physical models across scales \cite{rasp2020coupled,brenowitz2020interpreting,chantry2021opportunities,balogh2021toy} (e.g., failure in \emph{a posteriori} tests despite performing well in the \emph{a priori} setting \cite{pawar2020priori,guan2022stable,guan2022learning,pawar2023frame}).
\end{enumerate}

For the first pitfall, operator networks appear as a viable solution as they learn the mappings between infinite-dimensional function spaces. Deep operator networks (DeepONet) \cite{lu2021learning} and Fourier neural operator (FNO) \cite{li2020fourier} are the most popular frameworks in this area nowadays, showing varying levels of success in a wide range of benchmark problems \cite{lu2022comprehensive}. However, the training of operator networks requires large amounts of high fidelity data, which is often hard to collect in practice. Multifidelity operator network (MFON) \cite{howard2022multifidelity} leverages the use of an array of data and physical models with different fidelities to learn accurate mappings between the input and output spaces. \emph{ In this work, we formulate the closure problem (equivalently the subgrid scale correction and physical parameterization problem) as a multifidelity operator learning problem}. In particular, the simulation model resolving the large scales (the GPOD model in this case) represents the low fidelity model and the objective is to learn the corresponding high-fidelity correction terms to maintain accurate (and stable) predictions for the POD expansion coefficients. In the present study, we adopt the multifidelity implementation of the DeepONet architecture proposed by \citet{howard2022multifidelity}.

To address the second issue, differentiable programming tools provide the computing support to train the ML model in conjunction with other routines/solvers that interact with it. This mode of training is referred to as ``in-the-loop'' training \cite{um2020solver} and has resulted in significant accuracy and stability improvement in coupled physics-ML models \cite{kochkov2021machine,thuerey2021pbdl,list2022learned}. \emph{We extend the ``in-the-loop'' training paradigm to the closure modeling problem using MFONs}. Although this incurs a computational overhead for the gradient computations during the backpropagation, we argue that exposing the MFON model to its own predictions during the training phase is beneficial for long forecast lead times. It is particularly important to understand (1) how the inaccuracies of the low-fidelity GPOD model feed back into the estimated correction (high-fidelity) terms, and (2) how the uncertainty in ML outputs at one step propagates through the GPOD model and the ML model itself during the following time steps. 

In  \cref{sec:problem}, we consider a class of dynamical systems governed by unsteady partial differential equations (PDEs) along with their PROM formulation. We also show the need for the correction term to account for the modal truncation (coarsening) of the system. In \cref{sec:mfon}, we describe the MFON framework and its adaptation for the closure modeling problem in PROMs. Then, we dedicate \cref{sec:train} to differentiate between the traditional ``offline'' training of multifidelity DeepONets and its ``in-the-loop'' version. Numerical experiments using prototypical flow problems are given in \cref{sec:result}. Finally, concluding remarks and ideas for future work are provided in \cref{sec:conclusion}.

\section{Problem formulation} \label{sec:problem}
We consider a family of dynamical systems governed by PDEs as follows:
\begin{equation}
    \dfrac{\partial u}{\partial t} = \mathcal{N}(u, \dfrac{\partial u}{\partial x}, \dfrac{\partial^2 u}{\partial x^2}, \dots ;\mu), \label{eq:pde}
\end{equation}
where $u(x,t): \mathbb{R}^d \times \mathbb{R} \to \mathbb{R}^r$ is the solution, $x\in \mathbb{R}^d$ is the spatial dimension (e.g., $d\in \{1,2,3\}$), $t\in \mathbb{R}$ is the time, and {$\mu \in \mathbb{R}^p$} denotes the model's parameters. \Cref{eq:pde} is often solved numerically by defining a grid and applying one of the classical methods for solving PDEs (e.g., finite difference methods, finite volume methods, finite element method, spectral methods, etc.). This results in a semi-discretized system of equations for the solution state vector $\mathbf{u} \in \mathbb{R}^N$ as follows:
\begin{equation}
    \dfrac{d \mathbf{u}}{d t} = F(\mathbf{u};\mu), \label{eq:fom}
\end{equation}
where $F: \mathbb{R}^N \times \mathbb{R}^p \to \mathbb{R}^N$ represents the system's dynamics and $N$ is the number of degrees of freedom (e.g., number of grid points). For complex systems, the number of grid points, $N$, required to accurately resolve the underlying dynamics is usually very large. The associated computational costs make it unfeasible to embed such methods in multi-query applications, where the system is solved repetitively at different times and/or parameter values. Therefore, alternative methods to efficiently \emph{approximate} the solution $\mathbf{u}$ are sought-after.

\subsection{Reduced order modeling} \label{sub:rom}
PROMs have gained popularity in physical science disciplines (e.g., fluid dynamics). These methods are based on the Galerkin ansatz where the solution is defined as a linear superposition of a finite set of basis functions as follows:
\begin{equation}
    \mathbf{u}(t) \approx \mathbf{u}^{\text{ROM}}(t):= \sum_{i=1}^{R} a_i(t) \phi_i, \label{eq:upod0}
\end{equation}
where $\{\phi_i\}_{i=1}^{R}$ denote the basis functions (or spatial modes) and $\{a_i\}_{i=1}^{R}$ are the accompanying modal coefficients. POD has been the main driver for defining optimal sets of basis functions over the last few decades. The POD algorithm starts with an ensemble of system's realizations at different times as follows:
\begin{equation}
    \mathcal{U}:= \{\mathbf{u}(t_1), \mathbf{u}(t_2), \dots \mathbf{u}(t_M) \},
\end{equation}
where $\mathbf{u}(t_n)$ is a representation of the solution state vector at time $t_n$ that can be obtained from experimental measurements and more commonly from the high fidelity solution of \cref{eq:fom}, denoted as full order model (FOM) hereafter.

For parametric systems, this ensemble can be enriched with snapshot data at different parameter values as follows:
\begin{equation}
    \mathcal{U}:= \{\mathcal{U}^{\mu_1}, \mathcal{U}^{\mu_2}, \dots, \mathcal{U}^{\mu_P} \}.
\end{equation}

\Cref{eq:upod0} can be further equipped with an affine transformation as follows:
\begin{equation}
    \mathbf{u}^{\text{ROM}}(t) = \bar{\mathbf{u}} + \sum_{i=1}^{R} a_i(t) \phi_i, \label{eq:upod}
\end{equation}
where $\bar{\mathbf{u}}$ is a reference mode, usually taken as the ensemble-average. Therefore, the ensemble of shifted snapshots $\tilde{\mathcal{U}}$ is constructed from $\widetilde{\mathbf{u}}(t) = \mathbf{u}(t) - \bar{\mathbf{u}}$. A correlation matrix can be defined using $\widetilde{\mathcal{U}}$ and an eigenvalue decomposition reveals hierarchically-sorted basis functions to approximate $\mathbf{u}(t)$. Equivalently, a singular value decomposition of $\tilde{\mathcal{U}}$ can be effected as follows:
\begin{equation}
    \tilde{\mathcal{U}} = \mathbf{U} \boldsymbol{\Sigma} \mathbf{V}\tran, 
\end{equation}
where $\mathbf{U}$ and $\mathbf{V}$ are the matrices of left and right singular vectors, respectively and $\boldsymbol{\Sigma}$ is a diagonal matrix of the associated singular values. The first $R$ columns of $\mathbf{U}$ are used to define the POD basis functions $\{\phi_i\}_{i=1}^{R}$. The \emph{true} POD coefficients, which correspond to the best low-rank representation of $\mathbf{u}(t)$, can be computed as follows:
\begin{equation}
    \mathbf{a}(t_n) = \mathbf{\Phi}\tran (\mathbf{u}(t_n) - \bar{\mathbf{u}}). \label{eq:atrue}
\end{equation}

However, \cref{eq:atrue} requires access to the FOM solution $\mathbf{u}$, which is not possible in practice when the solution is queried at different times/parameters. Thus, we need to derive a model for the time-varying modal coefficients, $\mathbf{a}(t) = [a_1(t), a_2(t), \dots, a_R(t)]\tran$ without solving \cref{eq:fom}. This can be obtained by the orthogonal projection of the FOM operators in \cref{eq:fom} onto the POD basis functions $\mathbf{\Phi} = [\phi_1, \phi_2, \dots, \phi_R]$, a process known as Galerkin projection. This results in a Galerkin POD (GPOD) model as follows:
\begin{equation}
    \dot{\mathbf{a}} = \mathbf{\Phi}\tran F(\bar{\mathbf{u}} + \mathbf{\Phi} \mathbf{a} ;\mu). \label{eq:gpod}
\end{equation}

The GPOD predictions can be computed by numerically integrating \cref{eq:gpod} (e.g., using Runge-Kutta methods). Thus, the sequence of GPOD model predictions can be written as
\begin{equation}
    \mathbf{a}\gpod(t_{n+1}) = G(\mathbf{a}\gpod(t_{n}); \mu), \qquad \forall n \ge 0, \label{eq:agpod}
\end{equation}
where $G(\cdot;\cdot)$ is the GPOD one-step mapping (flow map) from $t_{n}$ to $t_{n+1}:=t_n + \Delta t$.

\subsection{Closure modeling} \label{sub:closure}
Due to the truncation of the POD basis functions (i.e., $R\ll N$), the accuracy of \cref{eq:gpod} for representing the dynamics of the resolved POD modes can be compromised. Thus, the GPOD predictions often deviate significantly from the optimal values in \cref{eq:atrue} and closure models are required to reduce this gap. In general, closure terms can appear as a correction to the right-hand side of \cref{eq:gpod} at continuous-time level, which is also known as the memory. Instead, we follow a predictor-corrector approach to compensate for the effect of truncated scales as follows:
\begin{equation}
    \begin{aligned}
        \widehat{\mathbf{a}}(t_{n+1}) &= G(\mathbf{a}(t_{n}); \mu), \\
        \mathbf{a}(t_{n+1}) &=  \widehat{\mathbf{a}}(t_{n+1}) + \mathbf{c}(t_{n+1}).
    \end{aligned} \label{eq:closure}
\end{equation}

A substantial body of work in the literature has been devoted to model the closure term $\mathbf{c}$. ML approaches and particularly DNNs have been utilized to develop closure models from data. A recent survey of such methods can be found in \cite{ahmed2021closures}. However, DNNs' capabilities are largely restricted to learning functions, which requires re-training the neural network for new parameter values and/or initial (or boundary) conditions. Instead, operator learning methods have been recently proposed to allow the training/testing of neural networks with varying settings (e.g., different parameters) \cite{lu2021learning}. In the following section, we briefly describe the DeepONet architecture and its multifidelity extension. Then, we address the closure modeling problem as a multifidelity learning task.

\section{Multifidelity operator networks for closure modeling} \label{sec:mfon}
\citet{lu2021learning} proposed the DeepONet framework to learn operators between infinite-dimensional spaces. Inspired by the universal approximation theorem for operators by \citet{chen1995universal}, DeepONet approximates the action of an operator $\mathcal{Q}(\mathbf{g})(\mathbf{y})$ as the inner product of the outputs of two neural networks, namely the branch net and trunk net. Although variations of DeepONet have been presented in \cite{lu2021learning}, the \emph{unstacked} implementation is widely considered the ``standard'' DeepONet where the branch net takes the input function $\mathbf{g}$ sampled at points $\{{x}_i\}_{i=1}^{S}$ and produces the output $\{B_i\}_{i=1}^{L}$. On the other hand,  the coordinates and/or parameters, denoted as $\mathbf{y}$, go to the trunk net whose output is denoted as $\{T_i\}_{i=1}^{L}$. Thus, the branch and trunk nets are trained simultaneously and operator $\mathcal{Q}(\mathbf{g})(\mathbf{y})$ is approximated as follows:
\begin{equation}
    \mathcal{Q}(\mathbf{g})(\mathbf{y}) \approx \mathcal{Q}^{\Theta}(\mathbf{g})(\mathbf{y}) := \sum_{i=1}^{L} B_i T_i,
\end{equation}
where $\Theta$ represents the aggregated parameters (weights) of the branch and trunk nets. We note that the only restriction is that the input function $\mathbf{g}$ is sampled at the same locations $\{{x}_i\}_{i=1}^{S}$. However, there are no constraints on the output query points $\mathbf{y}$ nor the input $\mathbf{g}$ itself. To achieve such powerful approximation and generalization capabilities, DeepONet usually requires larger amounts of data sets and the training is more expensive than a typical DNN. Modified architectures and training algorithms have been proposed to mitigate this burden \cite{wang2022improved}, including the use of physics-based loss \cite{wang2021learning} similar to physics-informed neural networks (PINNs) \cite{raissi2019physics}.

Multifidelity learning algorithms leverage the existence of data and models with disparate levels of fidelities to build a framework that outperforms those who could be built using single fidelity data/models. In the context of DeepONet, two recent developments include the MFONs proposed by \citet{howard2022multifidelity} and \citet{lu2022multifidelity}, where bifidelity DeepONets are trained simultaneously using the high fidelity and low fidelity data sets. \citet{lu2022multifidelity} proposed various architectures to combine the low fidelity operator network (LFON) and the high fidelity operator network (HFON), including learning the residual of the LFON and using LFON prediction to augment the inputs of the HFON. On the other hand, \citet{howard2022multifidelity} proposed the use of three composite blocks, starting with the LFON, followed by a linear and nonlinear sub-networks to learn the linear and nonlinear correlations between low fidelity and high fidelity data. In addition, \citet{howard2022multifidelity} demonstrated that a non-composite MFON can benefit from low-fidelity physical models to replace the LFON block. Along similar lines, \citet{de2022bi} used DeepONet to learn the discrepancy between the true system's response and low fidelity model's predictions.

Inspired by the non-composite MFON setup from \cite{howard2022multifidelity} and the residual learning approach in \cite{lu2022multifidelity,de2022bi}, we view the PROM closure modeling in \cref{sub:closure} as a multifidelity learning problem. {We hypothesize that there is a correlation between the low fidelity model predictions (for the resolved scales) and the contribution of the truncated scales onto the resolved scales (the closure terms).} We use the GPOD model $G(\cdot;\cdot)$ to define the low fidelity physical model and HFON learns the residual terms to correct the PROM predictions. In particular, we use the POD coefficients at current time, $\mathbf{a}(t_{n})$, and the low-fidelity predictions, $G(\mathbf{a}(t_{n});\mu)$) as inputs to the branch net. {This also introduces a time-memory effect similar to the non-Markovian terms in the Mori-Zwanzig formalism.} The model's parameter $\mu$ and the mode index $k \in \{1,2,\dots,R\}$ are fed to the trunk net. Thus, the MFON approximation of the closure term can be written as follows:
\begin{equation}
      \mathbf{a}(t_{n+1}) = \underbrace{G(\mathbf{a}(t_{n}); \mu)}_{\text{low fidelity model}} + \underbrace{\mathcal{Q}^{\Theta}\overbrace{\bigg([\mathbf{a}(t_{n})\tran,G(\mathbf{a}(t_{n}); \mu)\tran]\tran\bigg)}^{\text{branch net}} \overbrace{\bigg([\mu,k]\tran\bigg)}^{\text{trunk net}}}_{\text{residual}}. \label{eq:mfon}
\end{equation}

{While we are not using time explicitly as an input variable to MFON in \cref{eq:mfon}, we are learning the closure values as a function of the modal coefficients which themselves depend on time. Thus, at different times, the predicted corrections are different. It is also worth noting that the proposed MFON framework is not restricted to GPOD as the low fidelity model. For example, \citet{grimberg2020stability} identified culprits in the traditional Galerkin framework and recommended the use of Petrov-Galerkin framework instead. In particular, the least-squares Petrov-Galerkin (LSPG) has shown outstanding performance in many challenging problems \cite{carlberg2011efficient,carlberg2017galerkin} and can be an viable choice for the low fidelity model in MFON.}

\section{Training of multifidelity operator networks} \label{sec:train}
In this section, we differentiate between two paradigms for training MFONs. ML-based emulators are often trained as a standalone component where the training data are sampled from high fidelity simulations or experiments. This mode of training is denoted as ``offline'' training and has been predominantly adopted in scientific ML (SciML) for computational cost considerations. However, recent advances in differentiable programming and automatic differentiation (AD) tools have led to another mode of training, called ``online'' training. In particular, AD allows the computation of the gradients of arbitrary models' outputs with respect to their inputs (or parameters) so that the gradient-based optimizer can backpropagate through both the ML and physical models in an end-to-end setup \cite{de2018end,strofer2021end}. For time-dependent problems, we refer to the online training as ``in-the-loop`` training where both the physical and ML models are embedded and coupled in the time-integration loop. 

\subsection{Offline training}
Given the POD basis functions $\boldsymbol{\Phi}$ and the reference field $\bar{\mathbf{u}}$, the GPOD model $G(\cdot;\cdot)$ can be constructed (examples can be found in \cref{sec:result}). Two consecutive snapshots at time $t_{n}$ and time $t_{n+1}$ can be used to compute the \emph{true} POD coefficients as follows:
\begin{equation}
    \begin{aligned}
        \mathbf{a}(t_n) &= \boldsymbol{\Phi}\tran (\mathbf{u}(t_n) - \bar{\mathbf{u}}), \\
        \mathbf{a}(t_{n+1}) &= \boldsymbol{\Phi}\tran (\mathbf{u}(t_{n+1}) - \bar{\mathbf{u}}).
    \end{aligned}
\end{equation}

Meanwhile, the \emph{low-fidelity} prediction at $t_{n+1}$, given the \emph{true} coefficients at time $t_n$ can be written as follows:
\begin{equation}
    \widehat{\mathbf{a}}(t_{n+1}) = G(\mathbf{a}(t_n);\mu),
\end{equation}
and the correction can be defined as $\mathbf{c}(t_{n+1})= \mathbf{a}(t_{n+1})-\widehat{\mathbf{a}}(t_{n+1})$. Therefore, the training samples can be formed from $\{\mathbf{a}(t_{n}),\mathbf{a}(t_{n+1})\}$, where the branch net takes 
$\mathbf{a}(t_{n})$ and $\widehat{\mathbf{a}}(t_{n+1}) = G(\mathbf{a}(t_n);\mu)$ as input and the trunk net is fed with the parameters $\mu$ and the mode index $\{k\}_{k=1}^{R}$. Finally, the output of MFON can be compared against $\mathbf{a}(t_{n+1})$ using the following $\ell_2$ loss function:
\begin{equation}
    l(t_n,t_{n+1}) =  \bigg\| \mathbf{a}(t_{n+1}) - \widehat{\mathbf{a}}(t_{n+1}) -\mathcal{Q}^{\Theta}\bigg([\mathbf{a}(t_{n})\tran,\widehat{\mathbf{a}}(t_{n+1})\tran]\tran\bigg)\bigg([\mu,k]\tran\bigg) \bigg\|_2^2,
\end{equation}
where the parameters $\Theta$ can be optimized using a (stochastic) gradient descent-based optimizer (e.g., Adam). This mode of \emph{fully} supervised learning is denoted as \emph{offline training}. The underlying assumption here, which is often overlooked, is that the \emph{true} coefficients are available at current time $t_n$. Therefore, the ML model aims to learn a single-step correction. The rationale behind that stems from the assumption that the ML-based correction would always steer the \emph{low fidelity} predictions to match the \emph{high fidelity} solution. However, this is not often the case  and the ML prediction can, at best, be considered an approximation of the true residual. Therefore, the sequence of the MFON predictions can be written as follows:
\begin{equation}
    \begin{aligned}
        \widehat{\mathbf{a}}\mfon(t_{n+1}) &= G(\mathbf{a}\mfon(t_{n}); \mu) \\
        \mathbf{a}\mfon(t_{n+1}) &=  \widehat{\mathbf{a}}\mfon(t_{n+1}) +  \widehat{\mathbf{c}}(t_{n+1}),
    \end{aligned}
\end{equation}
where $\widehat{\mathbf{c}}$ denotes the MFON \emph{estimate} of the closure term $\mathbf{c}$. This discrepancy between the training and deployment conditions of MFON might cause a severe accumulation of the error, eventually leading to significantly inaccurate predictions. In addition, several studies have reported unstable behavior of the ML predictions after they are placed into the operation cycles despite their superior performance when tested separately (e.g., for single time steps).


\subsection{In-the-loop training}
\begin{figure}[ht]
    \centering
     \includegraphics[width=\linewidth]{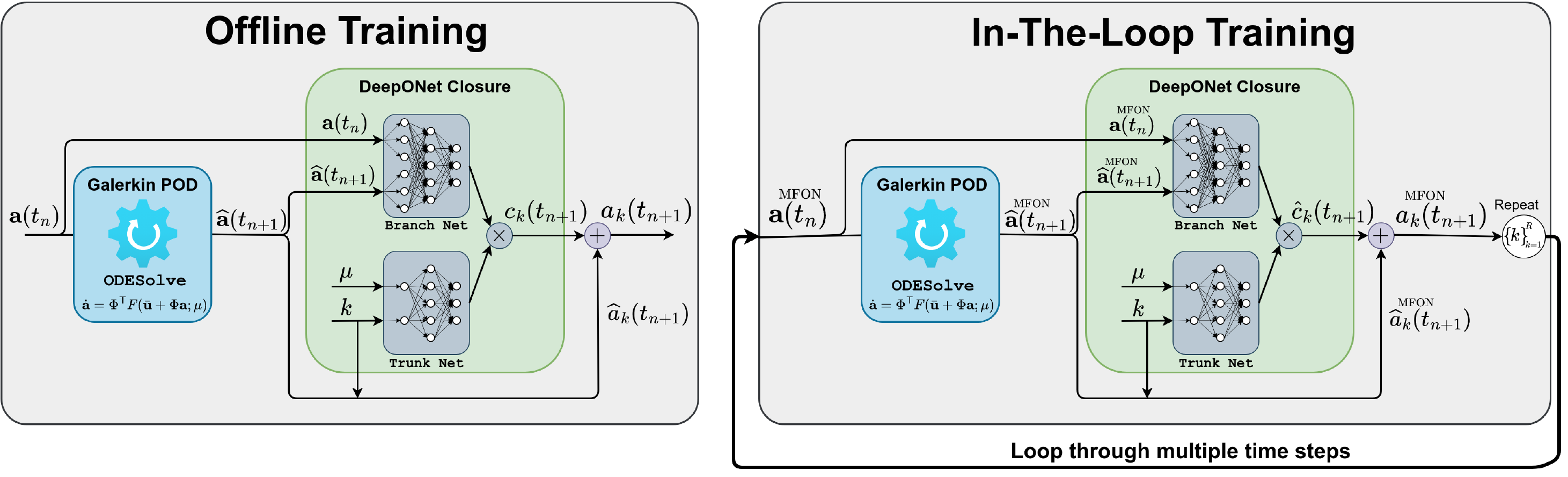}
     \vspace{-15pt}
    \caption{{A schematic diagram for conventional \emph{offline training} (left) and \emph{in-the-loop training} (right) of DeepONet in a multifidelity setting to correct Galerkin POD models predictions. For the offline training, the MFON is fed with \emph{true} values at the inputs. However, in-the-loop training allows the MFON to see the \emph{effect} of its previous output as input in the next step during the training process and thus accounts for the long-term interplay between the DeepONet and the GPOD model.}}
    \label{fig:loop}
\end{figure}

In order to create consistent training and testing environments, we leverage the differentiable programming capabilities of modern SciML tools (e.g., TensorFlow, PyTorch, and JAX) to create a feedback loop between the low fidelity and high fidelity models across multiple time steps. Instead of always feeding the MFON with \emph{true} data points at the current time, we use the predictions of MFON to define the input at next time step. This can be also thought of as a way to enforce \emph{temporal} causality. \Cref{fig:loop} shows a schematic representation comparing ``offline training'' where true coefficients are given at the input against ``in-the-loop training'' where MFON predictions at one time step are fed back as input for the next step. {Although in-the-loop training shares some features with recurrent neural networks, the key benefit of in-the-loop training lies in its combination with the physics-based solver (GPOD in our case) in a way that allows their interactions during the \emph{training} phase rather than using the solver only to pre-prepare the training data, e.g., through I/O operations.}

Ideally, MFON should be trained given only the initial condition at time $t_0$ and the output of each time step is looped back as input till the final time $t_M$. However, this is not often feasible due to the memory and compute costs of backpropagation algorithms as well as numerical precision errors resulting in exploding or vanishing gradients. Instead, we define a time window of $\tau$ time steps where the \emph{true} POD coefficients are only given at the first step after which MFON predictions are used recursively as follows:
\begin{equation}
    \begin{aligned}
        \widehat{\mathbf{a}}(t_{n+1}) &= G(\mathbf{a}(t_{n}); \mu), 
        \quad &
        \mathbf{a}\mfon(t_{n+1}) &=  \widehat{\mathbf{a}}(t_{n+1}) +  \widehat{\mathbf{c}}(t_{n+1}), \\
        \widehat{\mathbf{a}}\mfon(t_{n+2}) &= G(\mathbf{a}\mfon(t_{n+1}); \mu), 
        \quad &
        \mathbf{a}\mfon(t_{n+2}) &=  \widehat{\mathbf{a}}\mfon(t_{n+2}) +  \widehat{\mathbf{c}}(t_{n+2}), \\
        &\vdots \\
        \widehat{\mathbf{a}}\mfon(t_{n+\tau}) &= G(\mathbf{a}\mfon(t_{n+\tau-1}); \mu), 
        \quad &
        \mathbf{a}\mfon(t_{n+\tau}) &=  \widehat{\mathbf{a}}\mfon(t_{n+\tau}) +  \widehat{\mathbf{c}}(t_{n+\tau}). 
    \end{aligned} \label{eq:tau_seq}
\end{equation}
The resulting loss function can be written as follows:
\begin{equation}
    l(t_n,t_{n+\tau}) =  \dfrac{1}{\tau} \mathlarger{\sum}_{k=1}^{\tau} \bigg\| \mathbf{a}(t_{n+k}) - \widehat{\mathbf{a}}\mfon(t_{n+k}) -\mathcal{Q}^{\Theta}\bigg([\mathbf{a}\mfon(t_{n+k-1})\tran,\widehat{\mathbf{a}}\mfon(t_{n+k})\tran]\tran\bigg)\bigg([\mu,k]\tran\bigg) \bigg\|_2^2, \label{eq:tau_loss}
\end{equation}
where $\mathbf{a}\mfon(t_{n})=\mathbf{a}(t_{n})$ and $\widehat{\mathbf{a}}\mfon(t_{n+k})=\widehat{\mathbf{a}}(t_{n+k})$ as given in \cref{eq:tau_seq}. We note that the definition in \cref{eq:tau_loss} can be modified using a weighting scheme depending on the problem under consideration, e.g., to give higher importance to earlier predictions. In addition, the framework is extensible to cases where intermediate data points are not available for training MFON. For example, assuming that \emph{true} data are only available at the end of $\tau$ time steps, the loss function can be written as follows:
\begin{equation}
    l(t_n,t_{n+\tau}) =  \bigg\| \mathbf{a}(t_{n+\tau}) - \widehat{\mathbf{a}}\mfon(t_{n+\tau}) -\mathcal{Q}^{\Theta}\bigg([\mathbf{a}\mfon(t_{n+\tau-1})\tran,\widehat{\mathbf{a}}\mfon(t_{n+\tau})\tran]\tran\bigg)\bigg([\mu,k]\tran\bigg) \bigg\|_2^2. \label{eq:tau_loss_final}
\end{equation}

\begin{figure}[H]
    \centering
    \includegraphics[width=1.0\linewidth]{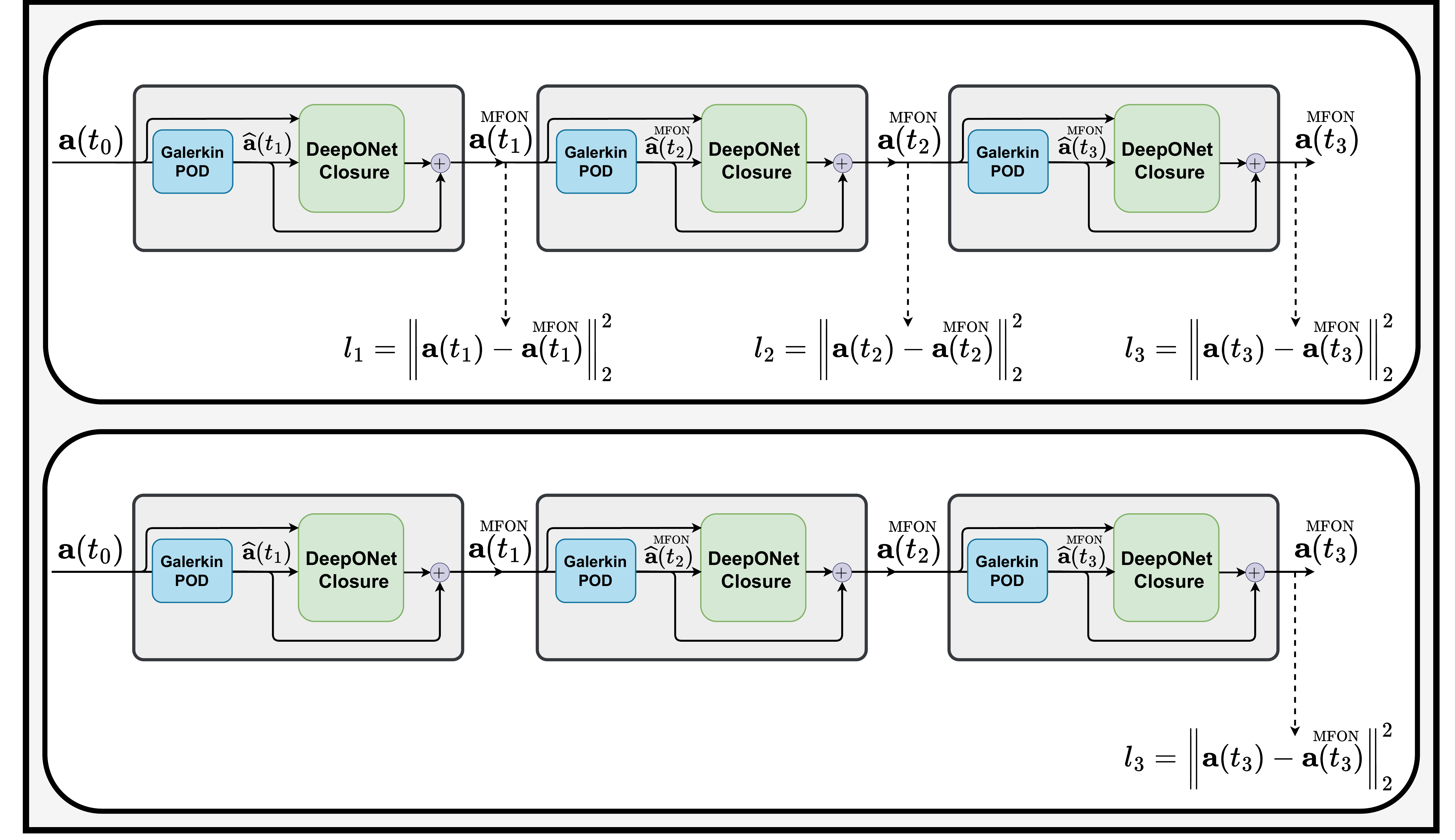}
    \vspace{-15pt}
    \caption{{An unrolled representation of Fig.~\ref{fig:loop} for \emph{in-the-loop training} of DeepONet over $\tau=3$ time steps. In top panel, the intermediate values of true POD coefficients are used to compute increments of the loss while only the final values after $\tau$ time steps are used to compute the loss in the bottom panel.}}
    \label{fig:unrolled}
\end{figure}

The unrolled representation of \emph{in-the-loop} training is shown in \cref{fig:unrolled} for $\tau=3$. {The idea of in-the-loop training for time-dependent problems can be also related to the recent windowed least-squares approaches (e.g., see \cite{parish2021windowed,shimizu2021windowed}), in the sense that the objective is to minimize the trajectory error rather than the instantaneous (single-step) error.} {However, in windowed PROM approaches, multiple (localized) models are often constructed for different \emph{non-overlapping} windows. During the prediction, these models are queried sequentially based on the corresponding point in space/time. In contrast, we build a single MFON model that is trained to minimize the solution error for $\tau$ steps and there is no restriction on where the time window begins. For example, a training batch can include \emph{overlapping} time windows as follows: $0\to \tau \Delta t$, $\Delta t\to (\tau+1) \Delta t$, $2\Delta t\to (\tau+2) \Delta t$, etc. Even though the MFON is trained to minimize the error for $\tau$ steps, having different overlapping trajectories in the training batches improves the generalizability of the trained MFON model. Therefore, during the prediction, the MFON model is called recursively as many time steps as needed to even extrapolate to times not seen during the training.} Although we only cover the MFON in discrete time setup with constant step size, it can be extended to continuous time to allow adaptive time stepping and sampling schemes.

{The value of the time window $\tau$ over which the MFON is trained is typically selected to balance two opposing effects. On one hand, $\tau$ should be as large as possible to impose higher temporal causality between the predictions of consecutive time steps. This minimizes the error along longer solution trajectories rather than just instantaneous (single-step) corrections. On the other hand, in practice, using large values of $\tau$ leads to lengthy computational graphs, which makes the computations of gradients (using automatic differentiation) more challenging. In addition, the selection of the optimizer, e.g., first order stochastic gradient descent-based optimizer, plays a role in the selection of $\tau$. Another important factor is related to the dynamical characteristics of the system itself. For example, for a chaotic system, it might be more challenging to use large values of $\tau$, in which cases the value of $\tau$ might be limited to a few Lyapunov times.}

\section{Numerical experiments} \label{sec:result}
We demonstrate the MFON framework for closure modeling in multiscale systems using two test problems showing strong convective dynamics. The first problem is the viscous Burgers problem corresponding to an advecting shock wave in one-dimensional (1D) setting. Then, we consider the two-dimensional (2D) vortex merger problem that has been used as a simplified model for many dynamical phenomena in large scale geophysical flows.

\subsection{Burgers problem} \label{sub:res-burgers}
The 1D viscous Burgers problem is defined using the following equation:
\begin{equation}
    \dfrac{\partial u}{\partial t} + u \dfrac{\partial u}{\partial x} = \dfrac{1}{\text{Re}} \dfrac{\partial^2 u}{\partial x^2}, \label{eq:brg}
\end{equation}
where $u(x,t): \mathbb{R} \times \mathbb{R} \to \mathbb{R}$ is the velocity field and $\text{Re}$ represents the Reynolds number (i.e., the ratio of inertial forces to viscous forces). The GPOD for \cref{eq:brg} can be written as follows:
\begin{equation}
\dot{\mathbf{a}} = \mathcal{C} + \mathcal{L} \mathbf{a} + \mathbf{a}\tran \mathcal{N} \mathbf{a}, \label{eq:gpodtensor}
\end{equation}
where $\mathcal{C}\in \mathbb{R}^{R}$, $\mathcal{L}\in \mathbb{R}^{R\times R}$, and $\mathcal{N}\in \mathbb{R}^{R\times R\times R}$ denote the constant, linear, and nonlinear terms as follows:
\begin{equation}
    \begin{aligned}
        \relax
        [\mathcal{C}]_{i} &= \bigg( \phi_i , -\bar{u} \dfrac{\partial \bar{u}}{\partial x} + \dfrac{1}{\text{Re}} \dfrac{\partial^2 \bar{u}}{\partial x^2} \bigg), \\
        [\mathcal{L}]_{ij} &= \bigg( \phi_i , -\bar{u} \dfrac{\partial \phi_j}{\partial x} -\phi_j \dfrac{\partial \bar{u}}{\partial x} + \dfrac{1}{\text{Re}} \dfrac{\partial^2 \phi_j}{\partial x^2}  \bigg), \\
        [\mathcal{N}]_{ijk} &= \bigg( \phi_i , -\phi_j \dfrac{\partial \phi_k}{\partial x} \bigg),
    \end{aligned}
\end{equation}
where $\big(\cdot,\cdot \big)$ denotes an inner product and $\bar{u}$ is the mean velocity field (see \cref{eq:upod}). The form of \cref{eq:gpodtensor} is often denoted as a tensorial GPOD and it takes advantage of the polynomial nonlinearity in governing equations to precompute the GPOD model terms. {As mentioned before, the proposed MFON framework is not restricted to a specific form of the low fidelity model and can easily incorporate cases with other types of nonlinearity. However, in those cases, hyper-reduction techniques \cite{cstefuanescu2014comparison,dimitriu2017comparative,carlberg2013gnat} should be implemented to reduce the computational cost of the low fidelity model.}

\begin{figure}[ht]
    \centering
    \includegraphics[width=\linewidth]{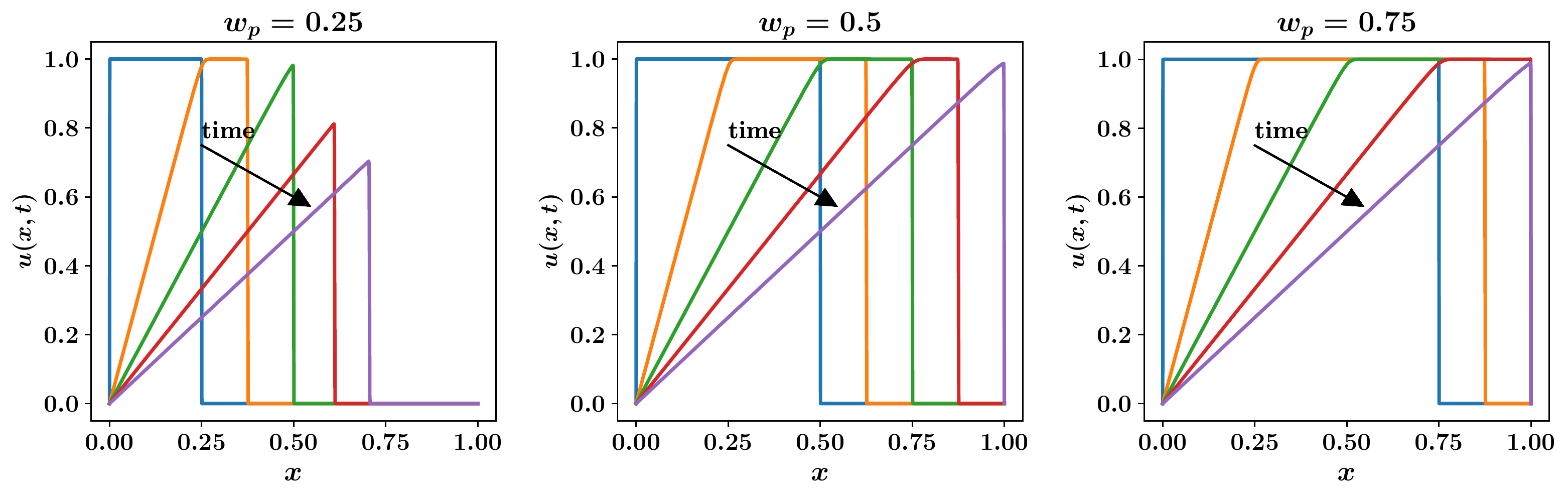}
    \vspace{-15pt}
    \caption{Velocity field at $\text{Re}=10,000$ starting with an initial condition of a pulse with a width $w_p \in \{0.25,0.50,0.75\}$}
    \label{fig:ufom}
\end{figure}

We consider a domain of length $1$ and impose zero Dirichlet boundary conditions (i.e., $u(0,t)=u(1,t)=0$) and define a family of initial conditions corresponding to a single square pulse of height $1$ and parameterized by the pulse width $w_p$ as follows:
\begin{equation}
    u(x,0) = \begin{cases} 
    1, \quad \text{if} \quad x \in [0,w_p],\\
    0, \quad \text{if} \quad x \in (w_p,1].
        \end{cases}
\end{equation}
In particular, our training data set corresponds to initial conditions with $w_p$ $\in [0.25, 0.75 ]$ with increment of $0.05$. Also, training data are generated at $\text{Re} \in [2500, 10000] $ with increment of $2500$.  For the FOM solution, we utilize a family of compact finite difference schemes for spatial discretization and the third order total variation diminishing Runge-Kutta (TVD-RK3) scheme for temporal integration. We divide the spatial domain into $4096$ equally spaced intervals and use a fixed time step of $\Delta t_{\text{FOM}} = 10^{-4}$. Snapshots of velocity field are stored every $100$ time steps for $t \in [0,1]$ to build the training data set. \Cref{fig:ufom} shows the time evolution of the resulting wave at $\text{Re}=10,000$ starting from three different pulse widths. For the GPOD models, we use a time step of $\Delta t_{\text{ROM}} = 10^{-2}$ that is $100$ times larger than $\Delta t_{\text{FOM}}$.

{For the design of the DeepONet architecture as well as selection of optimizer learning rate and number of iterations, we followed a manual trail-and-error approach since the dimensionality of the problem is relatively small. In particular, we use identical branch and trunk network architectures, except for the width of the input layer in each of them. For the branch net, the input layer width is $2R$ to accommodate the POD coefficients at the current time and the low fidelity predictions at the next time step. On the other hand, the width of the trunk input layer is 2, corresponding to Reynolds number and the mode index. Three hidden feedforward layers with $10$ neurons each are used for both the branch and trunk nets. For the activation function, we found that the hyperbolic tangent activation performs better than the rectified linear unit. For the optimizer, we use Adam with a decaying learning rate starting from $10^{-3}$.} \citet{howard2022multifidelity} include two DeepONets to learn the linear and nonlinear correlations between the low fidelity and high fidelity data. However, due to the \emph{in-the-loop} training framework, we find that the linear DeepONet causes the gradient descent optimizer to blow up due to the repetitive multiplications of weight matrices resulting in exploding gradients. Although such performance can be improved by enforcing stability conditions (e.g., using matrix decomposition \cite{drgona2021stochastic}), we only use the nonlinear DeepONet with the hyperbolic tangent activation function.

\subsubsection{Interpolative regime} \label{subsub:brg_inter}
\begin{figure}[ht]
    \centering
    \includegraphics[width=0.65\linewidth]{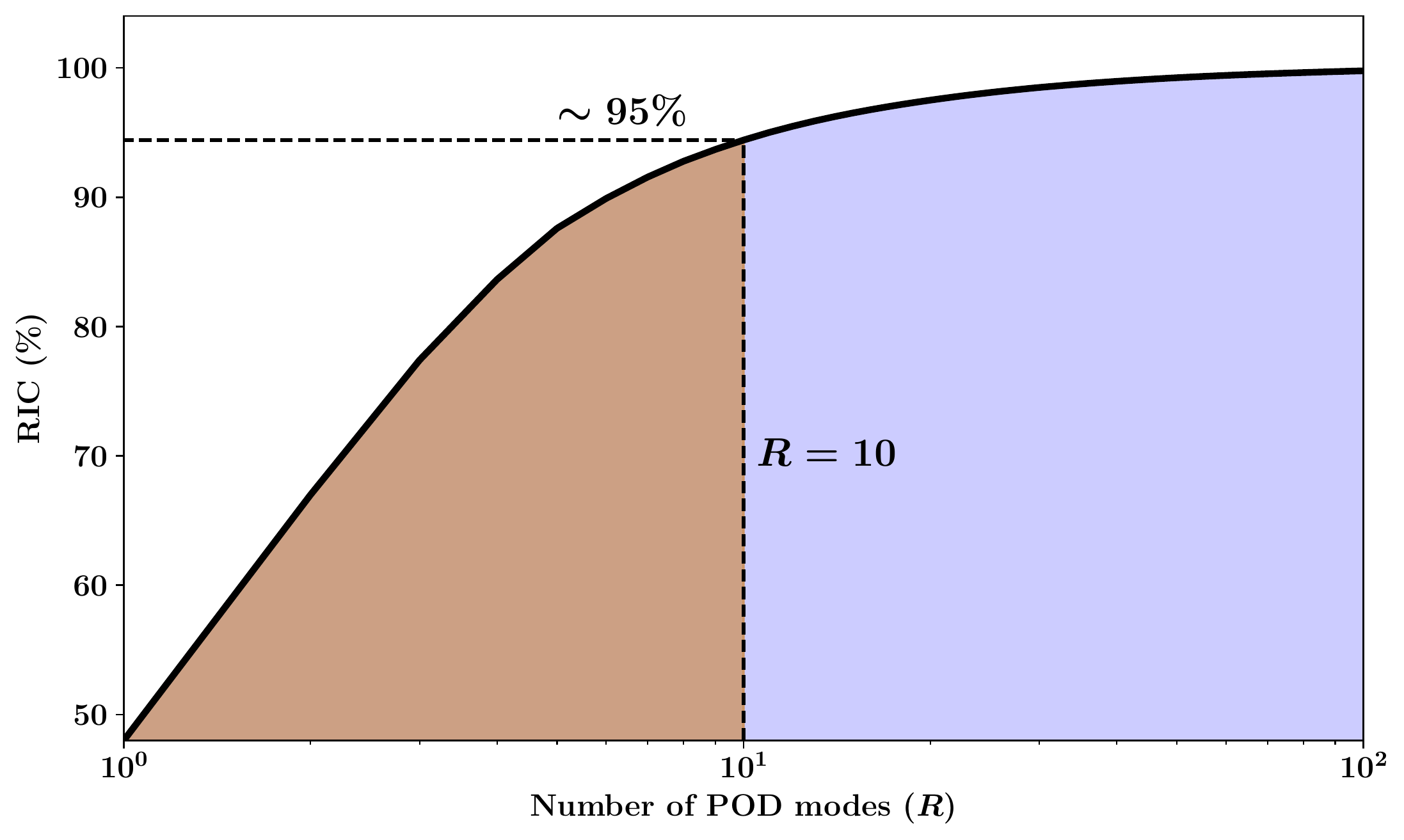}
    \vspace{-5pt}
    \caption{Relative information content for Burgers problem training data set.}
    \label{fig:uric}
\end{figure}

First, we test the MFON for closure modeling at parameter values that fall in the interpolation regime compared to the training data sets. In particular, we consider the Burgers problem with an initial condition defined by a pulse width of $0.675$ at Reynolds number of $4000$. We retain 10 modes to build the GPOD model, corresponding to $\sim 95\%$ of the total energy in the system quantified using the relative information content ($\text{RIC}$), as shown in \cref{fig:uric} and defined as follows:
\begin{equation}
    \text{RIC}(R) = \dfrac{\sum_{i=1}^{R} \sigma_i^2}{\sum_{i=1}^{M} \sigma_i^2}\times 100.
\end{equation}

\begin{figure}[H]
    \centering
    \includegraphics[width=\linewidth]{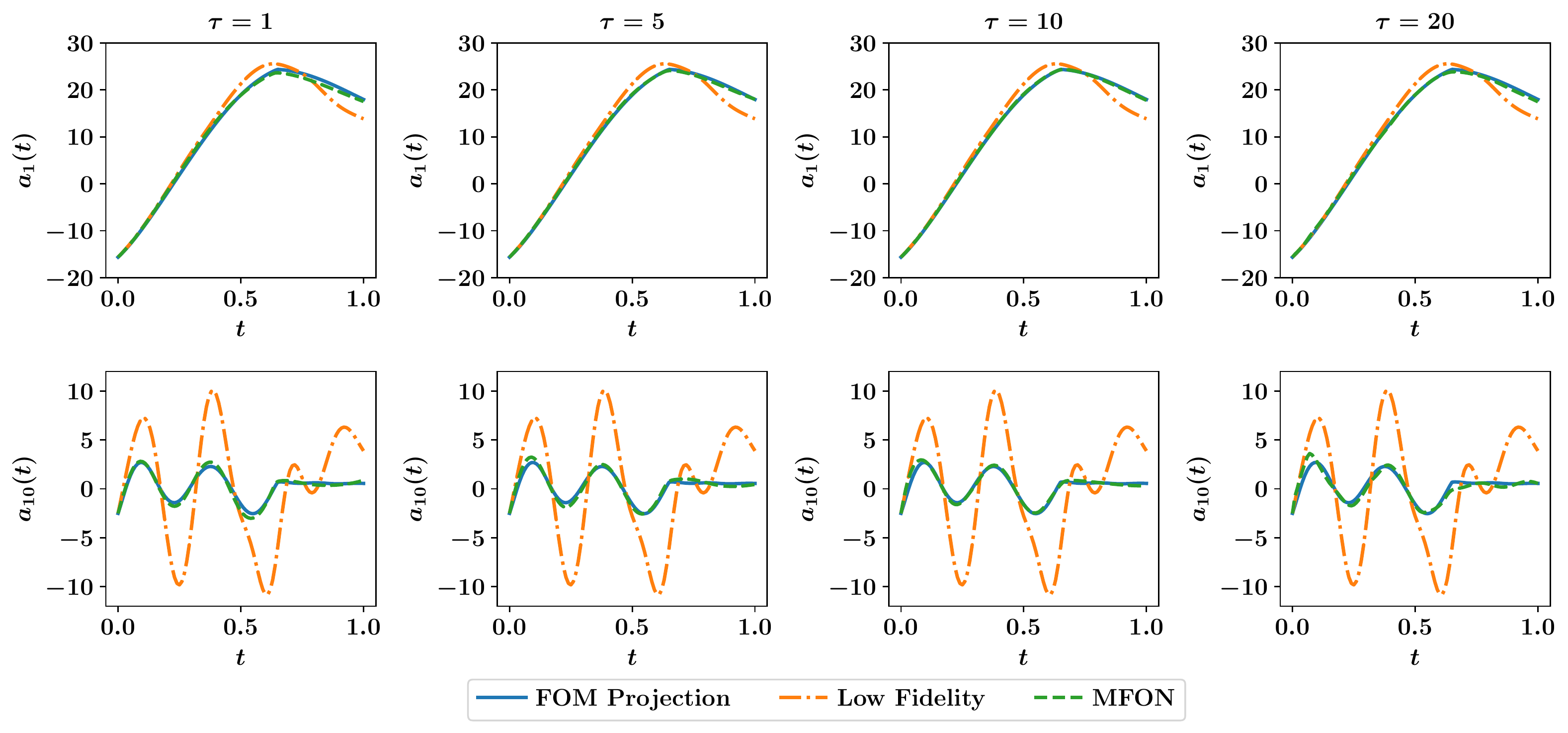}
    \vspace{-15pt}
    \caption{The evolution of the $1^{\text{st}}$ and $10^{\text{th}}$ POD modal coefficients for Burgers problem with an initial pulse width of $w_p=0.675$ at $\text{Re} = 4000$, corresponding to an interpolative test case. For in-the-loop training, true data points are assumed to be available at the end of $\tau$ time steps. Offline training is equivalent to setting $\tau=1$.}
    \label{fig:brg_coeff_interpolation}
\end{figure}

In addition, we compare the performance of MFON with offline training versus in-the-loop training. For the 1D Burgers problem, we found that both approaches of in-the-loop training (see \cref{fig:unrolled}) give similar results. Therefore, we consider the more challenging case where training data points are available only at the end of $\tau$ steps (i.e., the bottom panel in \cref{fig:unrolled}). \Cref{fig:brg_coeff_interpolation} depicts the predictions of the first and last POD coefficients. ``FOM Projection'' curves refer to the projection of the FOM snapshots onto the POD basis as defined in \cref{eq:atrue} while ``Low Fidelity'' refers to the predictions of the GPOD model without adding corrections as in \cref{eq:agpod}. It is clear the GPOD predictions deviate from their true values, especially for the high-index modes (e.g., $a_{10}$) that are closer to the truncated modes. This observation is consistent with the locality of modal interaction and energy transfer that has been motivating the development of variational multiscale closure techniques \cite{mou2021data}. The MFON results are shown for different values of $\tau$ that distinguishes the proposed in-the-loop training procedure. We also highlight that $\tau=1$ is equivalent to offline training as we shall refer to it from now on.

\begin{figure}[H]
    \centering
    \includegraphics[width=\linewidth]{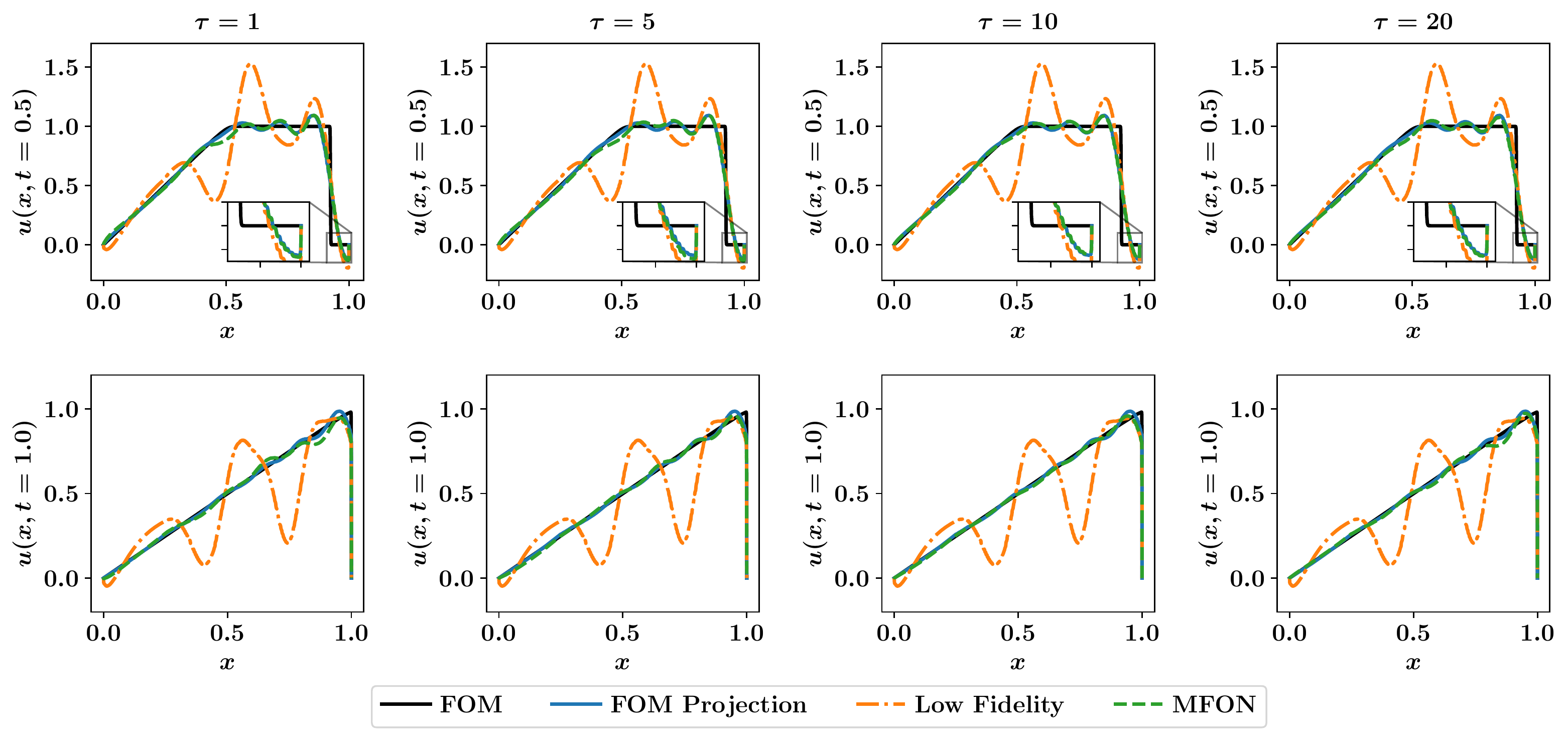}
    \vspace{-15pt}
    \caption{The predicted velocity field at $t=0.5$ (top) and $t=1.0$ (bottom) for Burgers problem with an initial pulse width of $w_p=0.675$ at $\text{Re} = 4000$, corresponding to an interpolative test case. For in-the-loop training, true data points are assumed to be available at the end of $\tau$ time steps. Offline training is equivalent to setting $\tau=1$.}
    \label{fig:brg_u_interpolation}
\end{figure}

\begin{figure}[H]
    \centering
    \includegraphics[width=\linewidth]{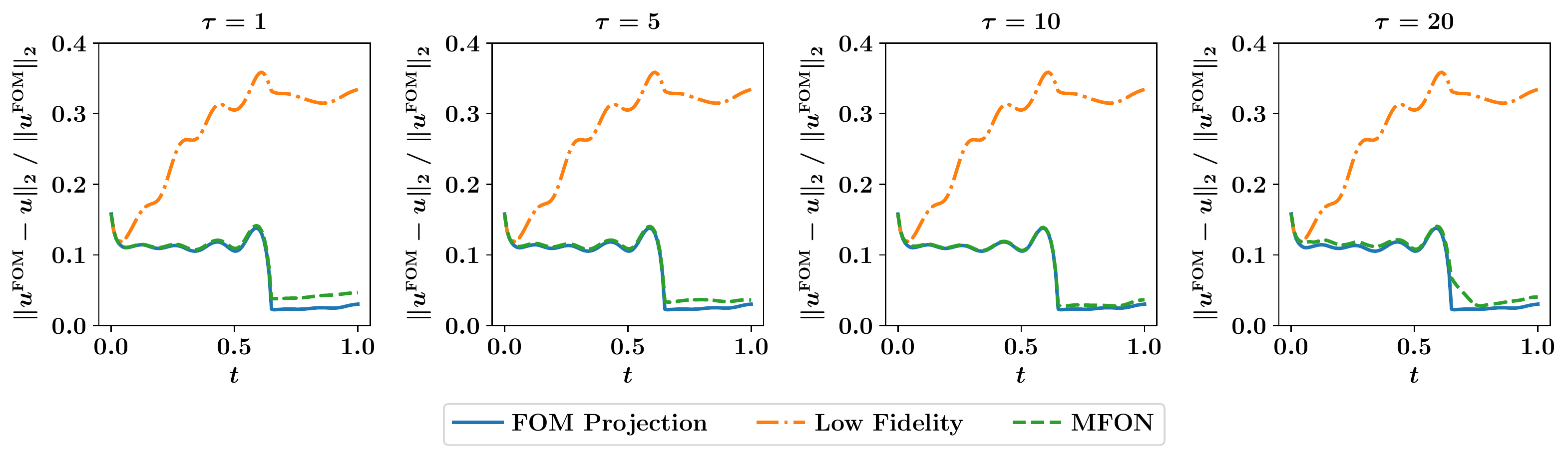}
    \vspace{-15pt}
    \caption{The relative error in the predicted velocity field as a function of time for Burgers problem with an initial pulse width of $w_p=0.675$ at $\text{Re} = 4000$, corresponding to an interpolative test case. For in-the-loop training, true data points are assumed to be available at the end of $\tau$ time steps. Offline training is equivalent to setting $\tau=1$.}
    \label{fig:brg_error_interpolation}
\end{figure}

\clearpage
\Cref{fig:brg_u_interpolation} shows that reconstructed velocity field at $t=0.5$ and $t=1.0$ by plugging the coefficients predicted by different models in \cref{eq:upod} while the FOM denotes the solution of the Burgers equation, \cref{eq:brg}, using finite difference schemes. The relative error with respect to $\mathbf{u}^{\text{FOM}}$ is presented in \cref{fig:brg_error_interpolation} where we can see that low fidelity GPOD predictions can lead to more than $30\%$ error in the velocity field. On the other hand, MFON accuracy levels are comparable to the FOM projection, which represents the \emph{maximum} reconstruction quality that can be obtained, for the retained POD modes. We also notice that \emph{in-the-loop} training yields slightly better MFON models compared to offline training, especially toward the end of the time interval.

\subsubsection{Extrapolative regime} \label{subsub:brg_extra}
\begin{figure}[ht]
    \centering
    \includegraphics[width=\linewidth]{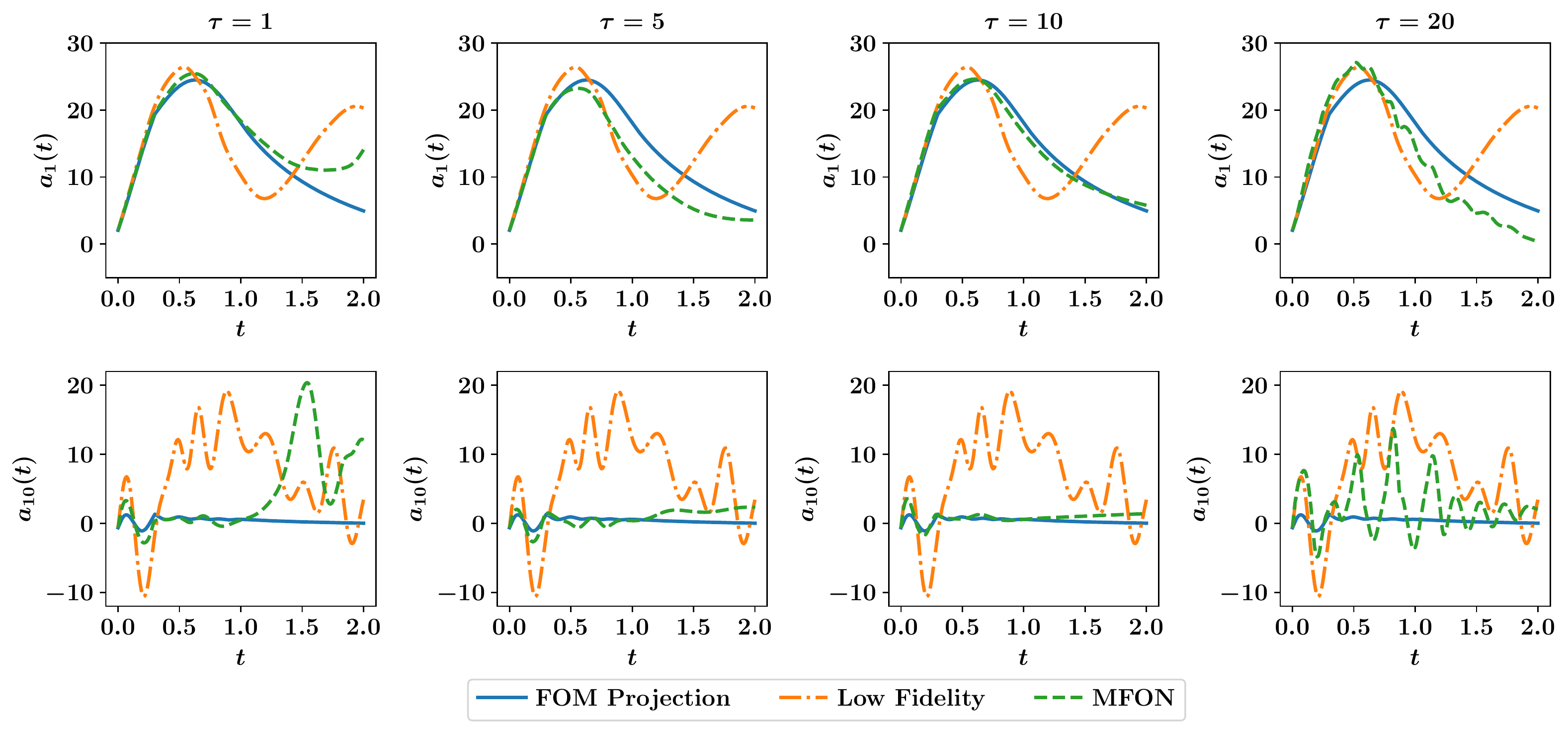}
    \vspace{-15pt}
    \caption{The evolution of the $1^{\text{st}}$ and $10^{\text{th}}$ POD modal coefficients up to $t=2$ for Burgers problem with an initial pulse width of $w_p=0.85$ at $\text{Re} = 15000$, corresponding to an extrapolative test case. For in-the-loop training, true data points are assumed to be available at the end of $\tau$ time steps. Offline training is equivalent to setting $\tau=1$.}
    \label{fig:brg_coeff_extrapolation}
\end{figure}

Next, we test the performance of MFON under extrapolation conditions. In particular, we use a Reynolds number of $15000$ that is $150\%$ higher than the largest value in the training data sets and we also consider a larger pulse width of $0.85$ to define the initial condition. More importantly, we explore the performance of MFON for larger time intervals. While the training data correspond to $t\in [0,1]$, we perform longer time predictions up to $t=2$. \Cref{fig:brg_coeff_extrapolation} shows the predictions of POD coefficients for the first and last mode. Although MFON with offline training performs well up to $t=1$, its accuracy significantly deteriorates for longer time predictions. On the other hand, MFON with in-the-loop training gives more accurate predictions with $\tau=5$ and $\tau=10$. However, $\tau=20$ results in significantly longer computational graphs for the backpropagation where exploding and diminishing gradient issues arise.

\begin{figure}[H]
    \centering
    \includegraphics[width=\linewidth]{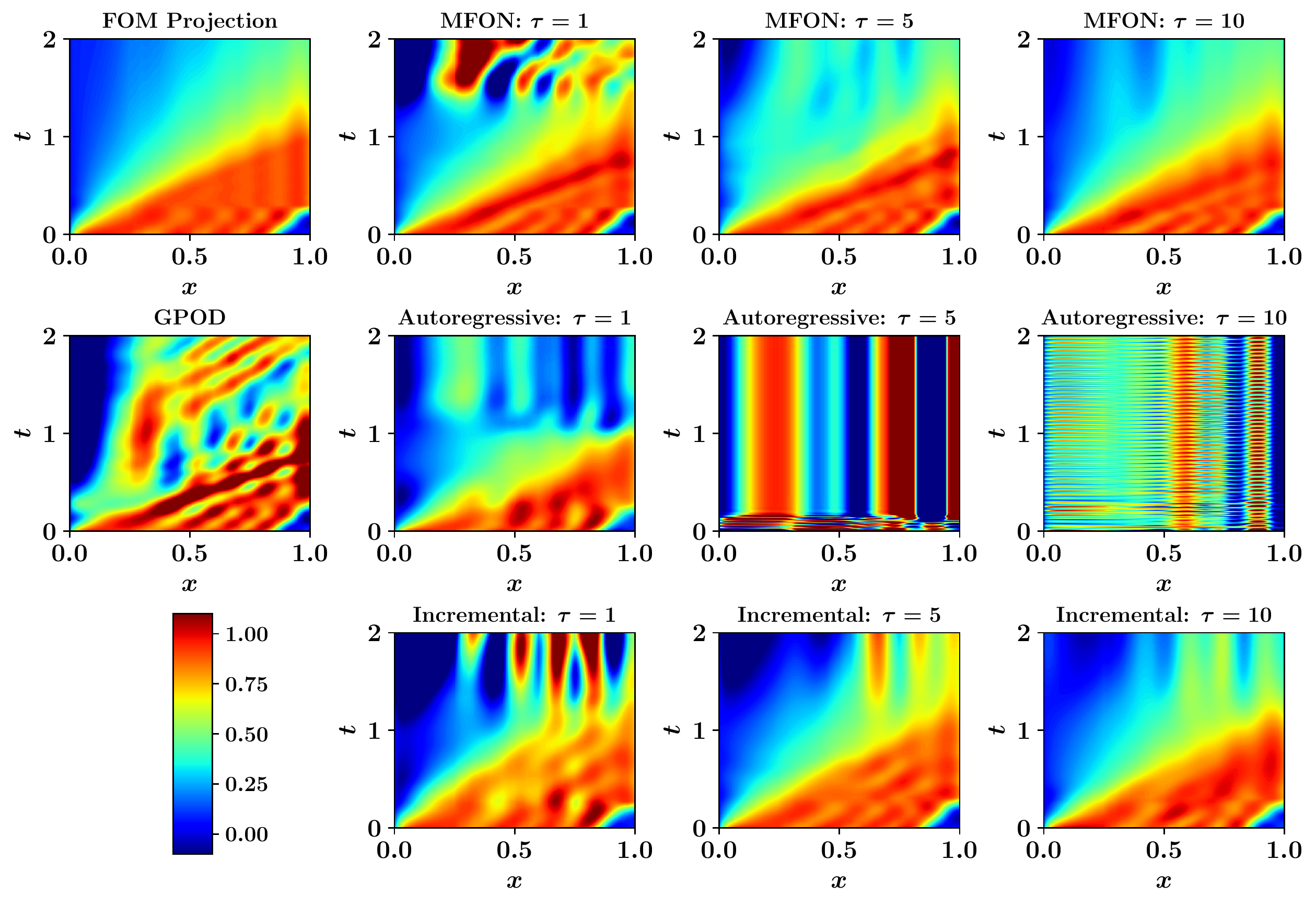}
    \vspace{-15pt}
    \caption{{The predicted velocity field for Burgers problem with an initial pulse width of $w_p=0.85$ at $\text{Re} = 15000$ using different operator learning approaches. For in-the-loop training, true data points are assumed to be available at the end of $\tau$ time steps. Offline training is equivalent to setting $\tau=1$.}}
    \label{fig:brg_u_extrapolation}
\end{figure}

\begin{figure}[H]
    \centering
    \includegraphics[width=\linewidth]{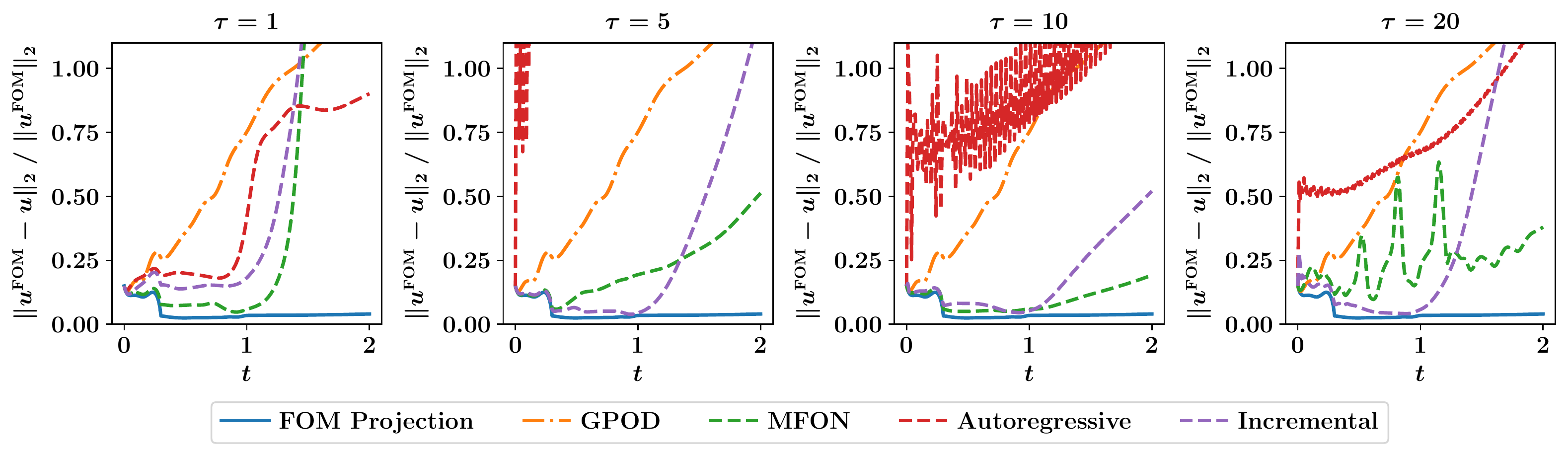}
    \vspace{-15pt}
    \caption{{The relative error in the predicted velocity field as a function of time for Burgers problem with an initial pulse width of $w_p=0.85$ at $\text{Re} = 15000$ with different modeling approaches. For in-the-loop training, true data points are assumed to be available at the end of $\tau$ time steps. Offline training is equivalent to setting $\tau=1$.}}
    \label{fig:brg_error_extrapolation}
\end{figure}

The space-time contour plots for the reconstructed velocity fields are shown in \cref{fig:brg_u_extrapolation} and the corresponding relative errors are given in \cref{fig:brg_error_extrapolation}. We see that GPOD can lead to around $100\%$ erroneous predictions. The accuracy levels of MFON with offline training are comparable to FOM projection up to $t=1$. Nonetheless, a drastic increase in the error occurs for temporal extrapolation. On the other hand, imposing temporal causality by in-the-loop training improves the predictability of MFON models with $\tau=5$ and $\tau=10$. It is also clear that there is an optimal value for $\tau$ with respect to the trade-off between the gain from introducing the feedback loop in the training and the cost of computing the gradients with longer computational graphs needed for backpropagation. We reiterate that the value of $\tau$ corresponds to the training phase only. During testing, it is assumed that only the initial condition at $t=0$ is known.

\begin{figure}[H]
    \centering
    \includegraphics[width=0.75\linewidth]{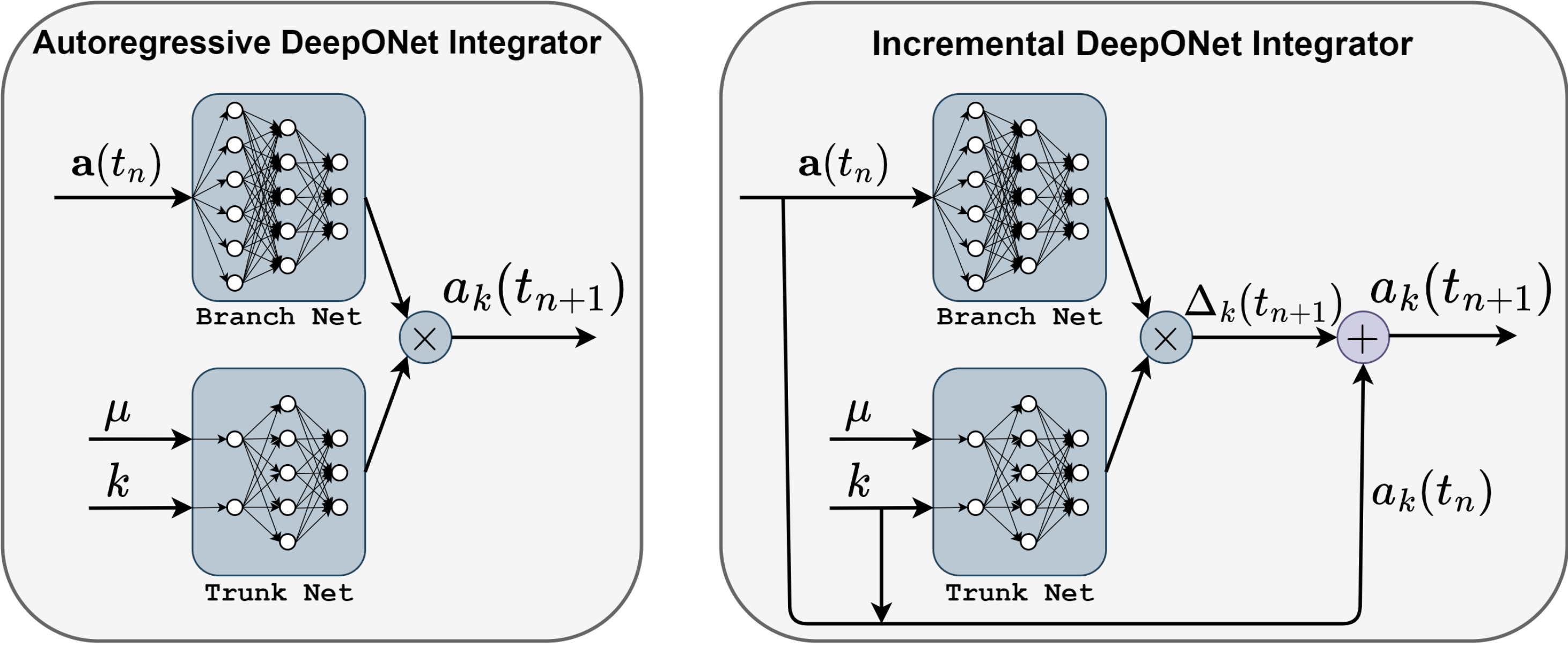}
    \caption{{Single fidelity DeepONet time integrators for the POD modal coefficients.}}
    \label{fig:single}
\end{figure}

{In addition, we compare the performance of MFON against a single-fidelity DeepONet to predict the evolution of the POD coefficients without taking any information from the low fidelity GPOD model. We consider two variations of such single-fidelity time integration model as shown in \cref{fig:single}. The first one corresponds to an autoregressive model that directly evolves the time-dependent coefficients as follows:
\begin{equation}
    a_k\ar(t_{n+1}) = \mathcal{Q}^{\Theta}\big(\mathbf{a}(t_{n})\big)\big([\mu,k]\tran\big).
\end{equation}
The predictions in \cref{fig:brg_u_extrapolation} and the corresponding errors in \cref{fig:brg_error_extrapolation} reveal that such single-fidelity (physics-agnostic) model does not perform well. In order to improve this performance, we consider an incremental DeepONet integrator as follows:
\begin{equation}
    a_k\inc(t_{n+1}) = a_k(t_{n}) + \mathcal{Q}^{\Theta}\big(\mathbf{a}(t_{n})\big)\big([\mu,k]\tran\big).
\end{equation}
\Cref{fig:brg_u_extrapolation,fig:brg_error_extrapolation} show substantial improvement in predictions using the incremental integrator compared to the autoregressive one. However, the performance drops beyond $t=1$, corresponding to extrapolation in time. We highlight that the incremental time integrator can be considered as a special case of multifidelity modeling, where the current state (at time $t_n$) represents the low fidelity estimate of the future state (at $t_{n+1}$). Nonetheless, using a physics-based GPOD as a low fidelity model in the MFON framework significantly improves the predictive capabilities in unseen regimes, compared to the fully data-driven models in \cref{fig:single}. }

\begin{figure}[H]
    \centering
    \includegraphics[width=\linewidth]{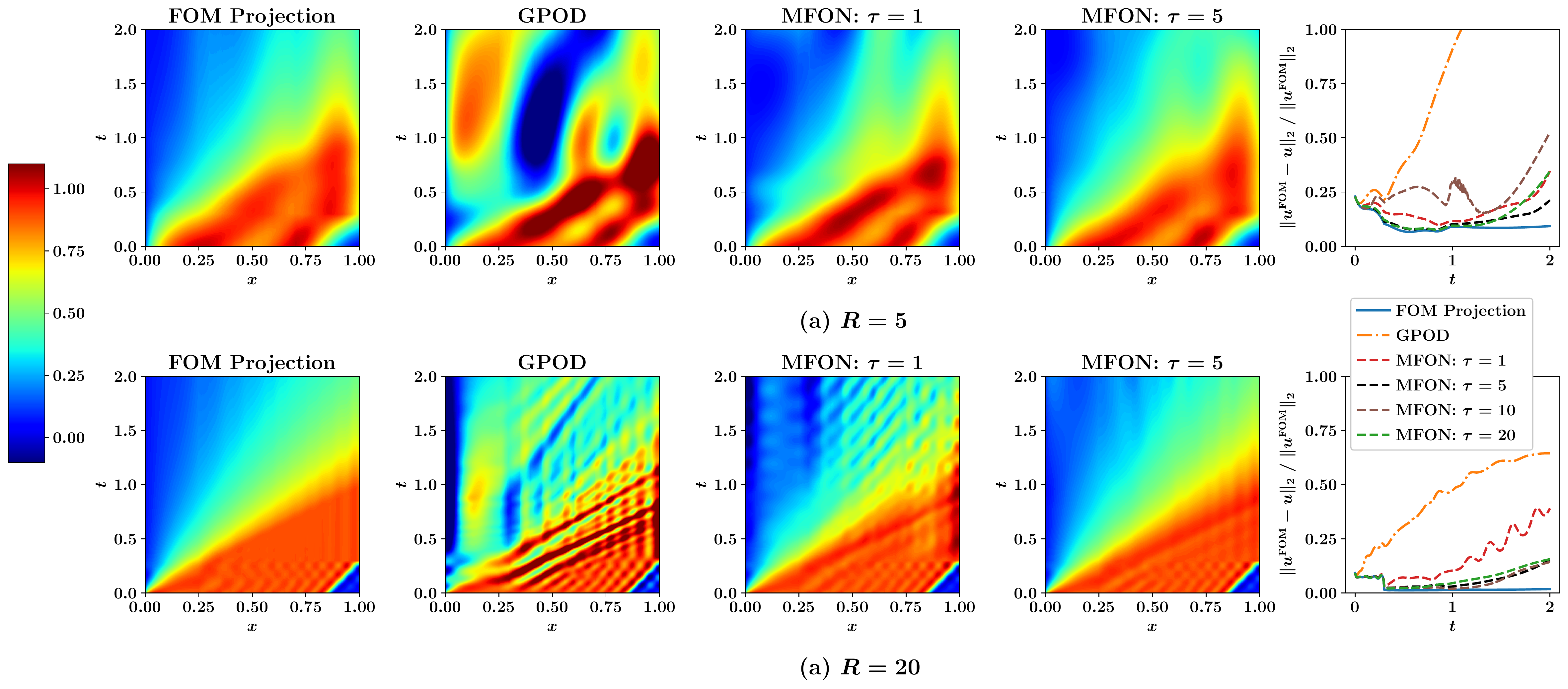}
    \vspace{-15pt}
    \caption{{The predicted velocity field and the relative error for Burgers problem with an initial pulse width of $w_p=0.85$ at $\text{Re} = 15000$, with different numbers of POD modes for the reduced order model.}}
    \label{fig:brg_nr}
\end{figure}

\clearpage
{Since MFON corrections are applied in the POD subspace, the lower bound for the MFON error corresponds to the projection error of the POD basis. In order to further reduce this error, either an increased number of POD basis should be retained in the PROM or an alternative set of basis functions should be adopted. For the latter, adaptive, localized, or custom-made basis functions can provide viable alternative, but it is outside the scope of the presents study. In order to assess the effect of varying the number of POD modes, \cref{fig:brg_nr} depicts the performance of MFON with $R=5$ and $R=20$, compared to the baseline of $R=10$ in the current study. We observe the MFON error is close to the FOM projection in different cases, implying a reduction in the closure error compared to the low fidelity GPOD predictions with the same number of POD modes. On the other hand, the upper bound for the MFON error relies on several factors. First, the specific architecture (e.g., number of layers, neurons, and activation functions) in addition to the optimizer options play a significant role. In this regard, there have been recent theoretical studies for the error estimates of DeepONet \cite{lanthaler2022error}, which is the key component in MFON. However, the extension of such analysis to multifidelity DeepONet settings is still missing. Second, the accumulation of time-integration error between different time steps plays another role, which depends on the order of the time integration scheme.}

\clearpage

\begin{figure}[H]
    \centering
    \includegraphics[width=0.6\linewidth]{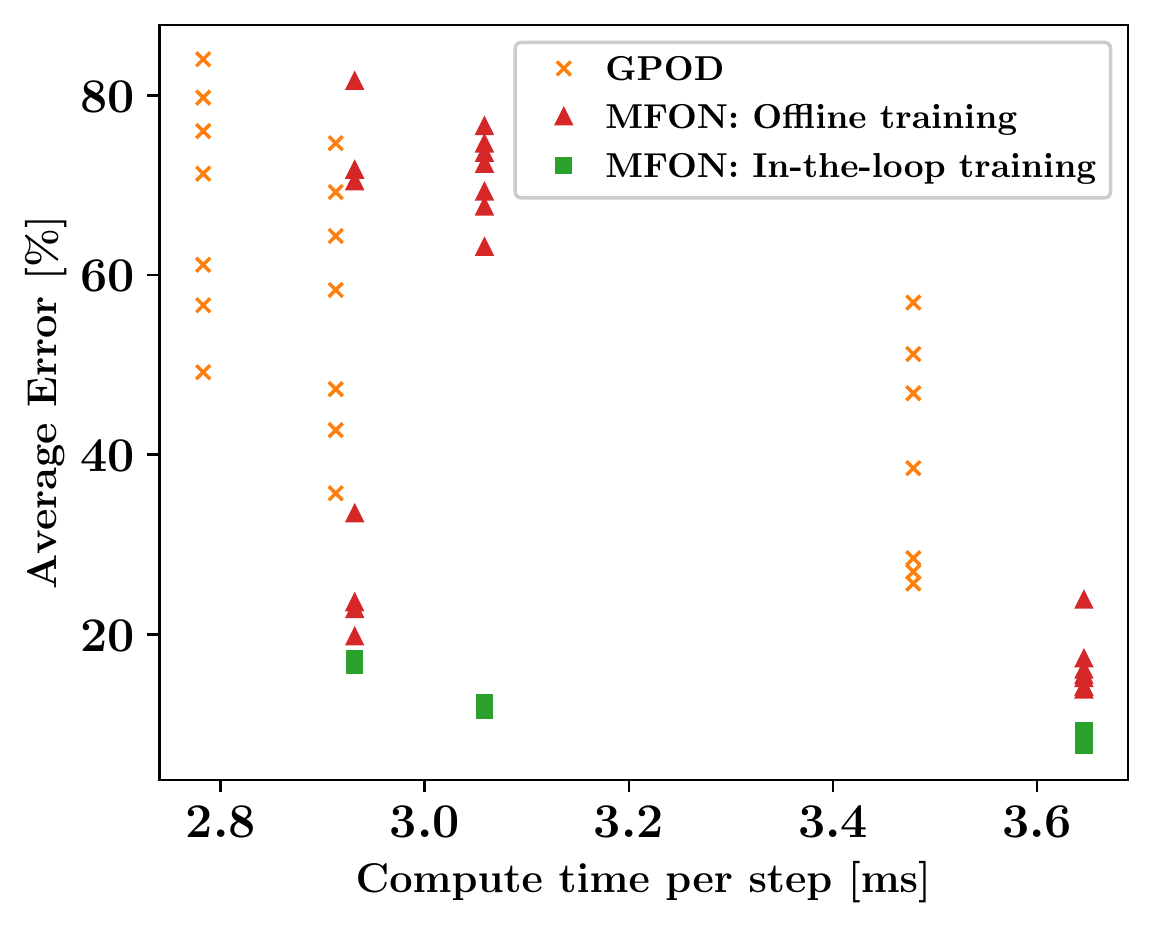}
    \vspace{-15pt}
    \caption{{Pareto front plot for the online compute time per step versus the time-averaged relative error in the reconstructed velocity field from different ROM approaches. Different data points (with the same color and marker) refer to testing at different Reynolds numbers. Points with the same compute time (vertical lines) correspond to fixed number of POD modes for each case, starting from $R=5$ on the left, $R=10$ in the middle, and $R=20$ on the right.}}
    \label{fig:brg_pareto}
\end{figure}

{The computational cost of MFON steps compared to GPOD is reported in \cref{fig:brg_pareto} for different test cases at different values of Reynolds number and different numbers of POD modes. We find that the computational overhead (in terms of compute time) of MFON over GPOD is about $5\%$, which is very minimal. We have benefited from JAX capabilities, including just-in-time (JIT) compilation and GPU porting to accelerate both the GPOD solver and the MFON computations.}

\subsection{Vortex merger problem} \label{sub:res-vortex}
Our second test problem is the 2D vortex merger problem, describing the evolution of two co-rotating vortices, initially in close proximity and eventually forming one big vortex. It models some of the fundamental processes in fluid motion that occur in many fields such as astrophysics,
meteorology, and geophysics. We consider a spatial domain of dimensions $(2\pi \times 2\pi)$ with periodic boundary conditions in both the $x$ and $y$ directions. The flow is initiated with a pair of Gaussian vortices with equal strengths centered at $(x_1,y_1)$ and $(x_2,y_2)$ as follows:
\begin{equation}
    \omega(x,y,0) =  \exp\left( -\rho \left[ (x-x_1)^2  + (y-y_1)^2 \right] \right) + \exp{\left( -\rho \left[ (x-x_2)^2 + (y-y_2)^2 \right] \right)}, \label{eq:vminit}
\end{equation}
where $\omega$ is the vorticity field and $\rho$ is a parameter that controls the mutual interactions between the two vortical motions, set as $\rho = \pi$ in the present study. We consider the two vortices' axes to be initially placed on a circle of radius $\pi/4$, separated by an $180^{\circ}$ angle as follows:
\begin{equation}
    \begin{aligned}
        x_1 &= \pi + \dfrac{\pi}{4}\cos(\theta),
        & \quad
        y_1 &= \pi + \dfrac{\pi}{4}\sin(\theta) \\
        x_2 &= \pi + \dfrac{\pi}{4}\cos(\theta+180^{\circ}),
        & \quad
        y_2 &= \pi + \dfrac{\pi}{4}\sin(\theta+180^{\circ}).   
    \end{aligned}
\end{equation}

The dynamics of the vortex merger problem can be described by the following vorticity transport equation:
\begin{equation}
\dfrac{\partial \omega}{\partial t} + J(\omega,\psi) = \dfrac{1}{\text{Re}} \nabla^2 \omega, \label{eq:vmerger}
\end{equation}
where $\psi$ denotes the streamfunction field that is linked with the vorticity field by the following kinematic relationship:
\begin{equation}
\nabla^2  \psi = -\omega. \label{eq:poisson}
\end{equation}
The Jacobian operator, $J(\cdot,\cdot)$, is defined as:
\begin{align}
    J(\omega,\psi) &= \dfrac{\partial \omega}{\partial x} \dfrac{\partial \psi}{\partial y} -  \dfrac{\partial \omega}{\partial y} \dfrac{\partial \psi}{\partial x}.
\end{align}

\begin{figure}[ht]
    \centering
    \includegraphics[width=\linewidth]{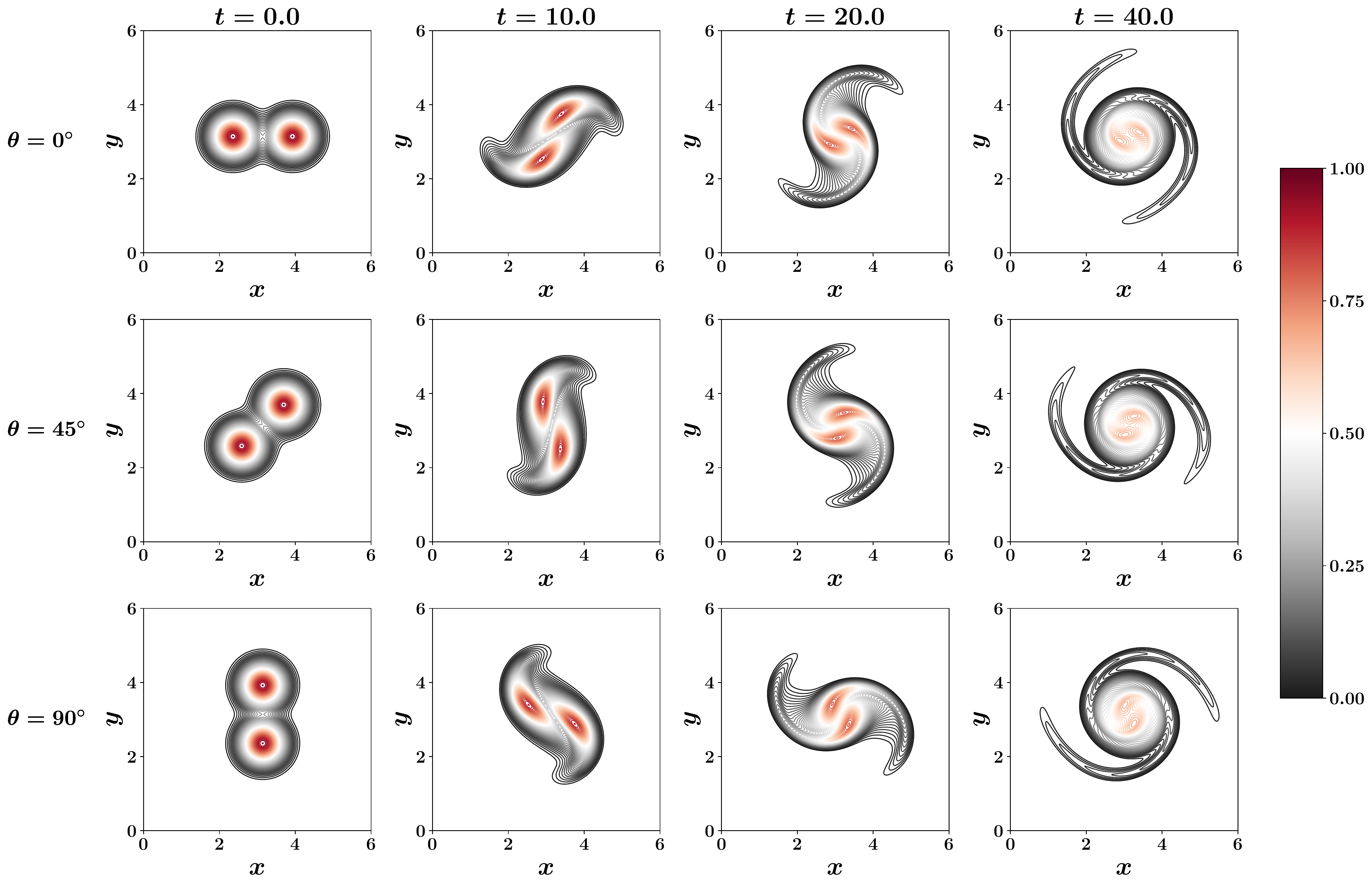}
    \caption{{The evolution of the vorticity field for the vortex merger problem at $\text{Re} = 2000$ starting from different initial conditions.}}
    \label{fig:wfom}
\end{figure}

For the FOM simulations, we define a regular Cartesian grid with a resolution of $256\times256$ (i.e., $\Delta x = \Delta y = 2\pi/256$) and we use the TVD-RK3 scheme with a time-step of $10^{-3}$. We run the FOM up to $t=40$. However, training data comprise vorticity snapshots that are collected every $100$ time steps only for $t\in [0,20]$. The evolution of the vortex merger problem is depicted in~\cref{fig:wfom}, which illustrates the convective and interactive mechanisms affecting the transport and development of the two vortices. This makes it a challenging problem for standard ROM approaches and a good testbed for the proposed MFON framework.

We apply the POD analysis in \cref{sub:rom} onto the vorticity field since it is the prognostic variable in \cref{eq:vmerger}. Similar to \cref{eq:upod}, we approximate $\omega$ using the span of the first $R$ POD modes as follows:
\begin{equation}
    \omega(\cdot,t) \approx  \omega^{\text{ROM}}(\cdot,t) =  \bar{\omega}(\cdot) +  \sum_{i=1}^R a_i(t) \phi_i(\cdot), \label{eq:wpod}
\end{equation}
On the other hand, the streamfunction can be approximated as follows:
\begin{equation}
    \psi(\cdot,t) \approx  \psi^{\text{ROM}}(\cdot,t) =  \bar{\psi}(\cdot) +  \sum_{i=1}^R a_i(t) \theta_i(\cdot), \label{eq:spod}
\end{equation}
where the mean field and basis functions for the streamfunction can be obtained using \cref{eq:poisson} as follows:
\begin{equation}
    \begin{aligned}
        \nabla^2 \bar{\psi} &= - \bar{\omega}, \\
        \nabla^2 \theta_i &= -\phi_i, \quad i=1,2,\dots, R. 
    \end{aligned} \label{eq:basis_lap}
\end{equation}

It should be noted that enforcing the kinematic relationship in \cref{eq:poisson} does not guarantee that the resulting basis functions $\theta$ for the streamfunction are orthogonal. However, it allows us to use the  same coefficients $\{a_i\}_{i=1}^{R}$ in \cref{eq:wpod} and \cref{eq:spod}. We set $R=10$ to define the total number of resolved scales and hence the dimensionality of the GPOD system. The GPOD for \cref{eq:vmerger} is similar to \cref{eq:gpodtensor} with the following terms:
\begin{equation}
    \begin{aligned}
        \relax
        [\mathcal{C}]_{i} &= \bigg( \phi_i , -J(\bar{\omega},\bar{\psi}) + \dfrac{1}{\text{Re}} \nabla^2 \bar{\omega} \bigg), \\
        [\mathcal{L}]_{ij} &= \bigg( \phi_i , -J(\bar{\omega},\theta_j)  -J(\phi_j,\bar{\psi}) + \dfrac{1}{\text{Re}} \nabla^2 \phi_j  \bigg), \\
        [\mathcal{N}]_{ijk} &= \bigg( \phi_i , -J(\phi_j,\theta_k) \bigg).
    \end{aligned}
\end{equation}

\subsubsection{Interpolative regime} \label{subsub:vortex_inter}
\begin{figure}[ht]
    \centering
    \includegraphics[width=\linewidth]{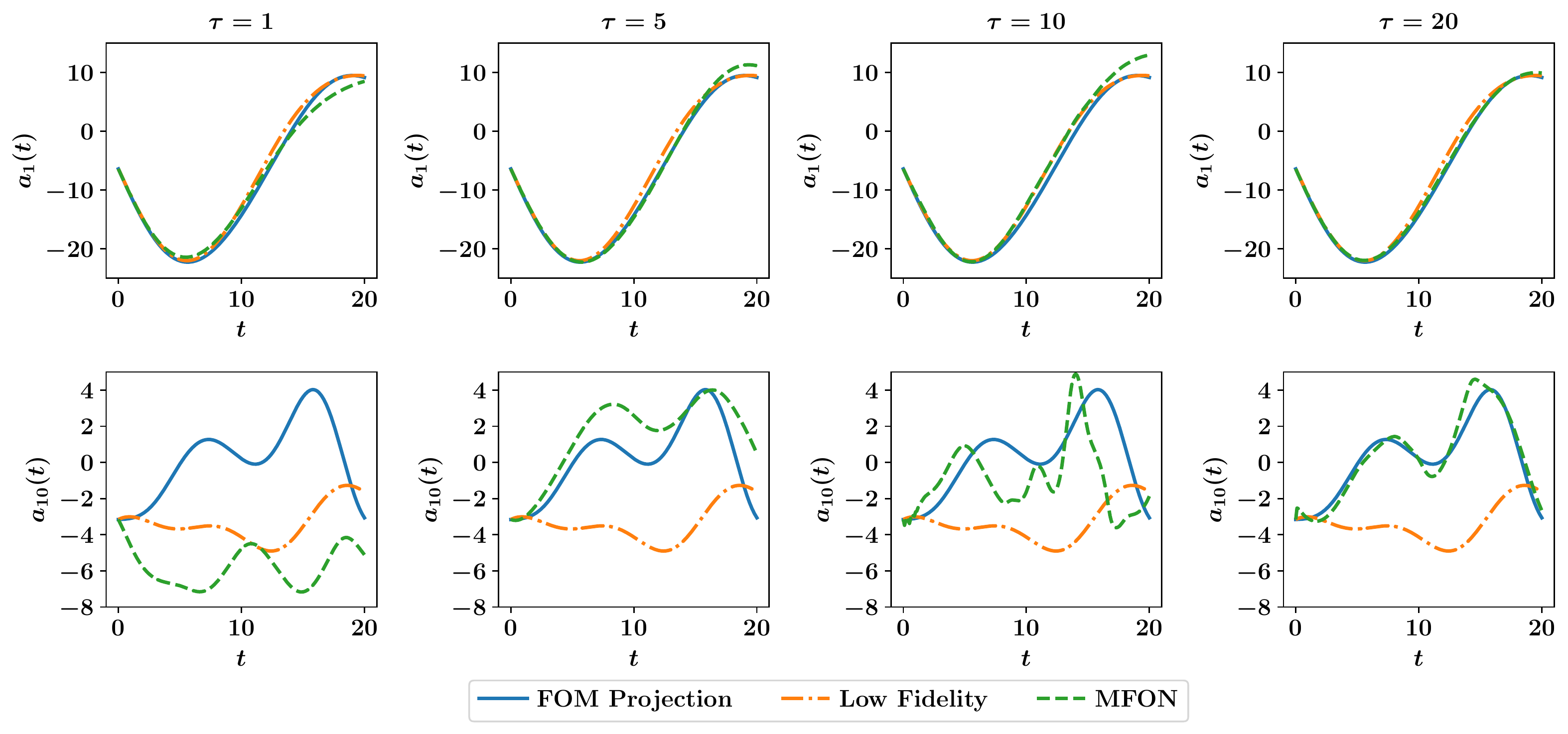}
    \caption{The evolution of the $1^{\text{st}}$ and $10^{\text{th}}$ POD modal coefficients up to $t=20$ for the vortex merger problem with an initial orientation with $\theta=60^{\circ}$ at $\text{Re} = 1500$, corresponding to an interpolative test case. For in-the-loop training, true data points are assumed to be available at the end of $\tau$ time steps. Offline training is equivalent to setting $\tau=1$.}
    \label{fig:vortex_acoeff_interpolation}
\end{figure}

We first demonstrate the performance of the proposed MFON for an interpolative testing case in both the initial condition and Reynolds number. We consider an initial vorticity field corresponding to $\theta=60^{\circ}$ and $\text{Re}=1500$ and run the FOM, GPOD, and MFON models up to $t=20$. The evolution of the first and last POD coefficients is shown in \cref{fig:vortex_acoeff_interpolation}. The low fidelity predictions, resulting from the GPOD model, exhibit large deviation from the FOM projected values, especially for the $10^{\text{th}}$ mode. Meanwhile, using MFON with offline training ($\tau=1$) exacerbates the predictions despite giving high accuracy levels for single step predictions (not shown). On the other hand, adopting in-the-loop training ($\tau>1$) significantly improve the results. However, it is important to select $\tau$ values that enforce temporal causality while keeping the training feasible with the available data and optimizer. For instance, we compare the performance with $\tau \in \{1,5,10,20\}$ and we find that $\tau=20$ gives the best results while $\tau=10$ yields the worst predictions for this particular problem. We also highlight these observations correspond to having training data available at the end of $\tau$ steps (bottom panel in \cref{fig:unrolled}).

\begin{figure}[ht]
    \centering
    \includegraphics[width=\linewidth]{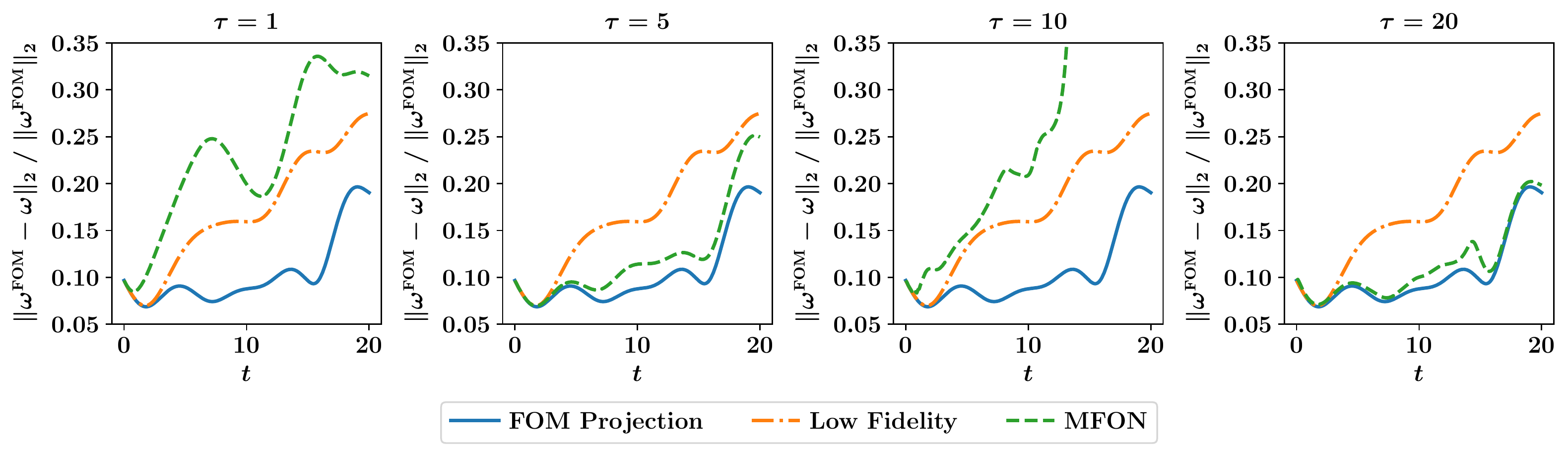}
    \caption{The relative error in the predicted vorticity field as a function of time for the vortex merger problem with an initial orientation with $\theta=60^{\circ}$ at $\text{Re} = 1500$, corresponding to an interpolative test case. For in-the-loop training, true data points are assumed to be available at the end of $\tau$ time steps. Offline training is equivalent to setting $\tau=1$.}
    \label{fig:vortex_error_interpolation}
    \end{figure}

\begin{figure}[ht]
    \centering
    \includegraphics[width=0.8\linewidth]{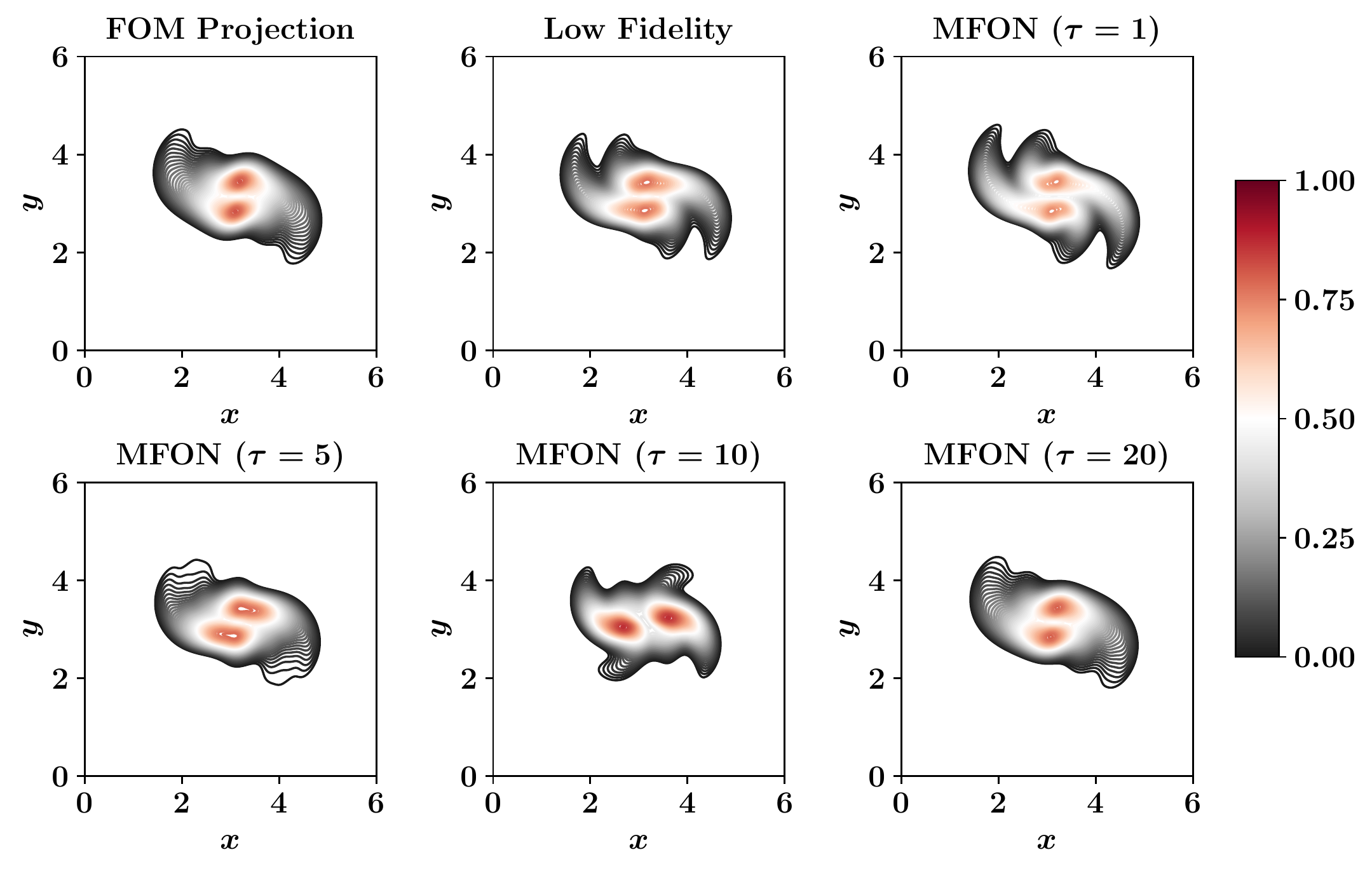}
    \caption{The predicted velocity field at $t=20$ for the vortex merger problem with an initial orientation with $\theta=60^{\circ}$ at $\text{Re} = 1500$, corresponding to an interpolative test case. For in-the-loop training, true data points are assumed to be available at the end of $\tau$ time steps. Offline training is equivalent to setting $\tau=1$.}
    \label{fig:vortex_vort_interpolation}
    \end{figure}

\Cref{fig:vortex_error_interpolation} displays the propagation of relative error in the predicted vorticity fields from various models. It is clear that offline training can yield unreliable predictions even for interpolative test cases. In addition, it is evident that the choice of $\tau$ is an important hyperparameter in our in-the-loop framework for training MFON. The reconstruction of vorticity field at $t=20$ (corresponding to the end of the training time interval) is shown in \cref{fig:vortex_vort_interpolation}. The MFON with $\tau=20$ is in close agreement with the FOM projection field, which defines the optimal reconstruction with $10$ POD modes. In addition, the time step size in GPOD and MFON is $100$ times larger than that of FOM. Therefore, we not only reduce the number of degrees of freedom from $256^2$ to $10$, but also use a much coarser time stepping.

\clearpage
\subsubsection{Extrapolative regime} \label{subsub:vortex_extra}
\begin{figure}[ht]
    \centering
    \includegraphics[width=\linewidth]{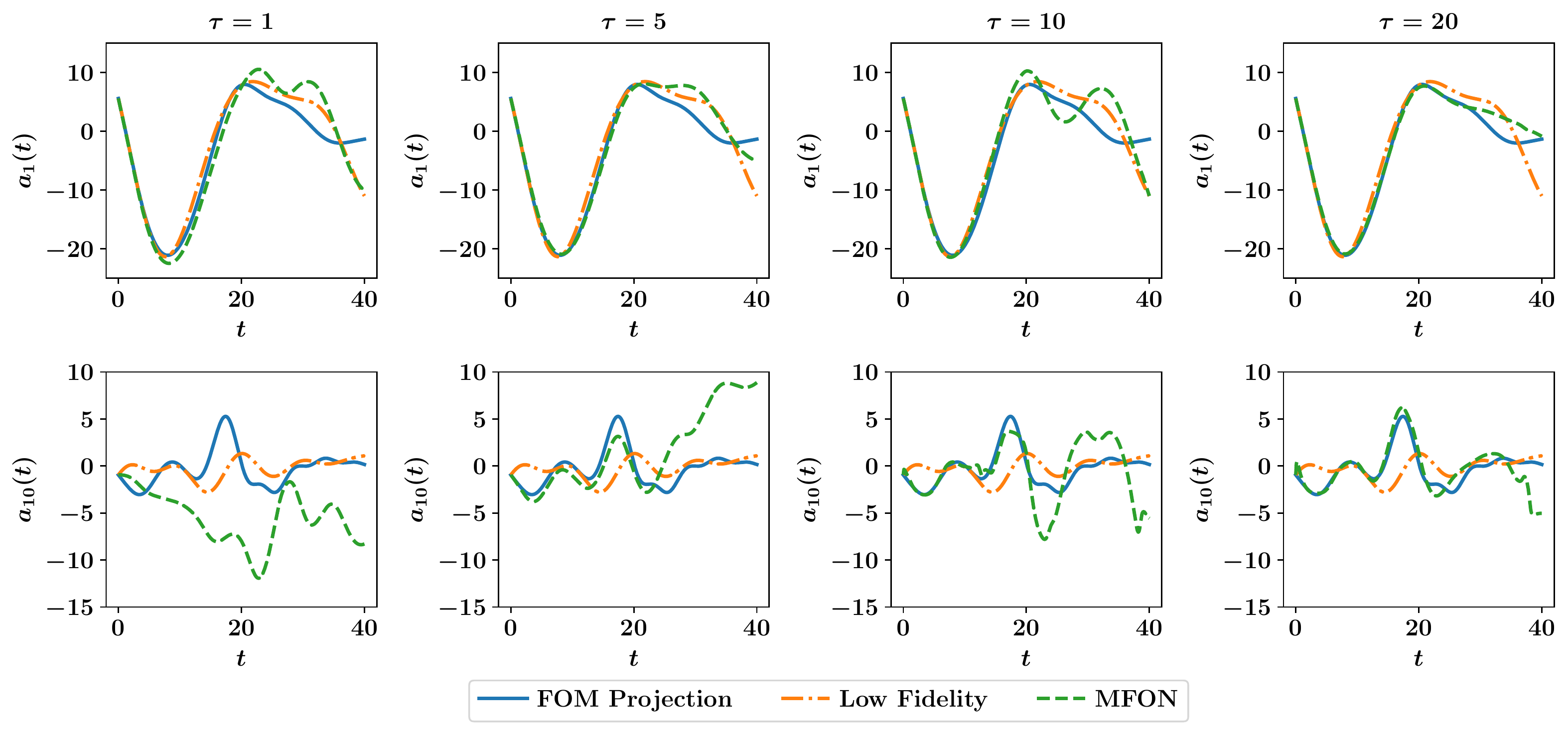}
    \caption{The evolution of the $1^{\text{st}}$ and $10^{\text{th}}$ POD modal coefficients up to $t=40$ for the vortex merger problem with an initial orientation with $\theta=45^{\circ}$ at $\text{Re} = 3000$, {corresponding to an extrapolative test case}. For in-the-loop training, true data points are assumed to be available at the end of $\tau$ time steps.}
    \label{fig:vortex_acoeff_extrapolation}
\end{figure}

\begin{figure}[ht]
    \centering
    \includegraphics[width=\linewidth]{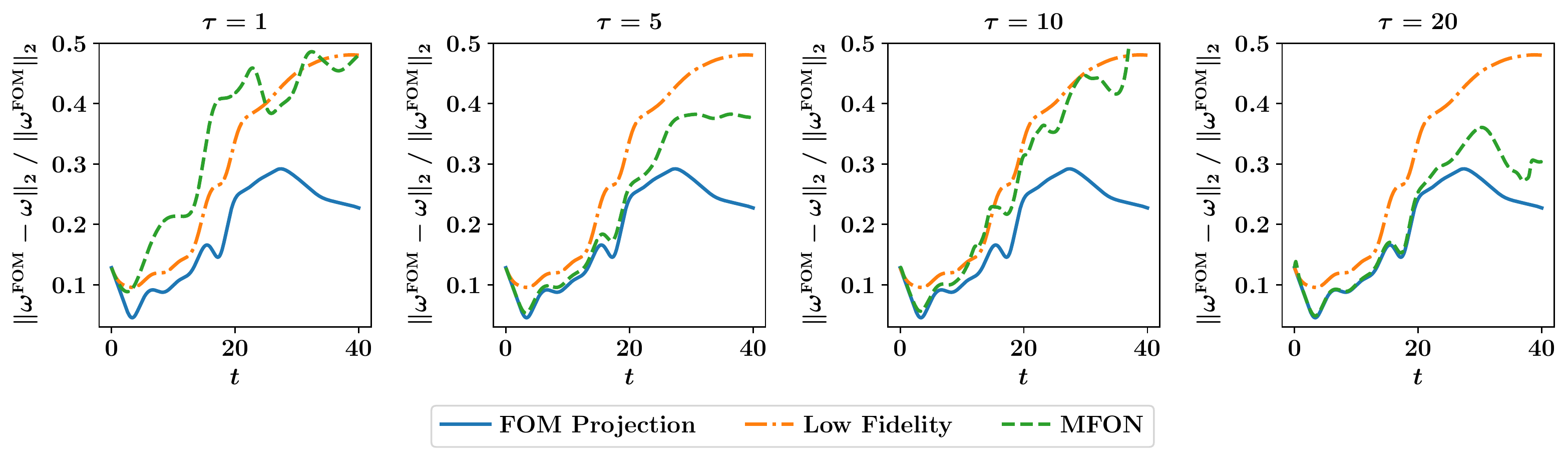}
    \caption{The relative error in the predicted vorticity field as a function of time for the vortex merger problem with an initial orientation with $\theta=45^{\circ}$ at $\text{Re} = 3000$, {corresponding to an extrapolative test case}. For in-the-loop training, true data points are assumed to be available at the end of $\tau$ time steps.}
    \label{fig:vortex_error_extrapolation}
\end{figure}

The extrapolative performance of MFON is corroborated for the vortex merger problem by considering $\text{Re}=3000$ that is $150\%$ larger than the highest value in the training data set. In addition, we carry out predictions up to $t=40$, which is twice the time interval in the training data. We emphasize that this is a transient and highly convective flow problem and is often challenging for ROMs. \Cref{fig:vortex_acoeff_extrapolation} shows the predictions of the $1^{\text{st}}$ and $10^{\text{th}}$ POD coefficients from $t=0$ to $t=40$ while the relative error in the reconstructed vorticity field is presented in \cref{fig:vortex_error_extrapolation}. The GPOD model without the closure term fails to capture the true dynamics of the resolved scales. Meanwhile, the MFON with offline training deviates significantly from the target solution. Introducing the feedback loop in the training environment (i.e., $\tau>1$) improves the results, especially for longer time predictions. We also notice that $\tau=10$ gives a worse model than both resulting from $\tau=5$ and $\tau=20$, similar to \cref{subsub:vortex_inter}.

\begin{figure}[ht]
    \centering
    \includegraphics[width=\linewidth]{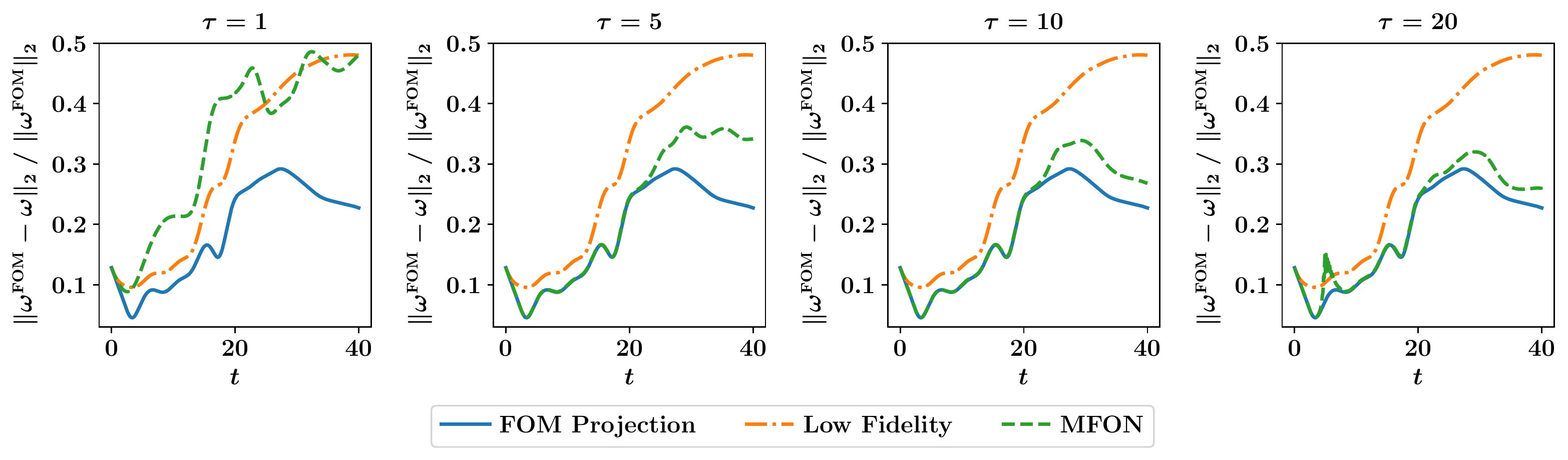}
    \caption{The relative error in the predicted vorticity field as a function of time for the vortex merger problem with an initial orientation with $\theta=45^{\circ}$ at $\text{Re} = 3000$, {corresponding to an extrapolative test case}. For in-the-loop training, true data points are assumed to be available also at the intermediate steps within the $\tau$ time steps.}
    \label{fig:vortex_error_extrapolation_more}
    \end{figure}

Finally, we illustrate the effect of using in-the-loop training while penalizing deviations at the intermediate points (top panel in \cref{fig:unrolled}). In particular, we suppose that training data are available at each time step as defined in \cref{eq:tau_loss}. However, this can be extended to arbitrary points that are unequally spaced in time. The relative error of the predicted vorticity field as a function of time is presented in \cref{fig:vortex_error_extrapolation_more}. We notice that the error of MFON with $\tau>1$ is about twice as accurate as the GPOD model without closure. Interestingly, the $\tau=10$ case is giving very good results, compared to \cref{fig:vortex_error_extrapolation} where the loss function is defined using data points at the end of $\tau$ steps. The choice of optimal values of $\tau$ and its dependence on the availability of intermediate data points is still an open research question.   

\begin{figure}[H]
    \centering
    \includegraphics[width=0.8\linewidth]{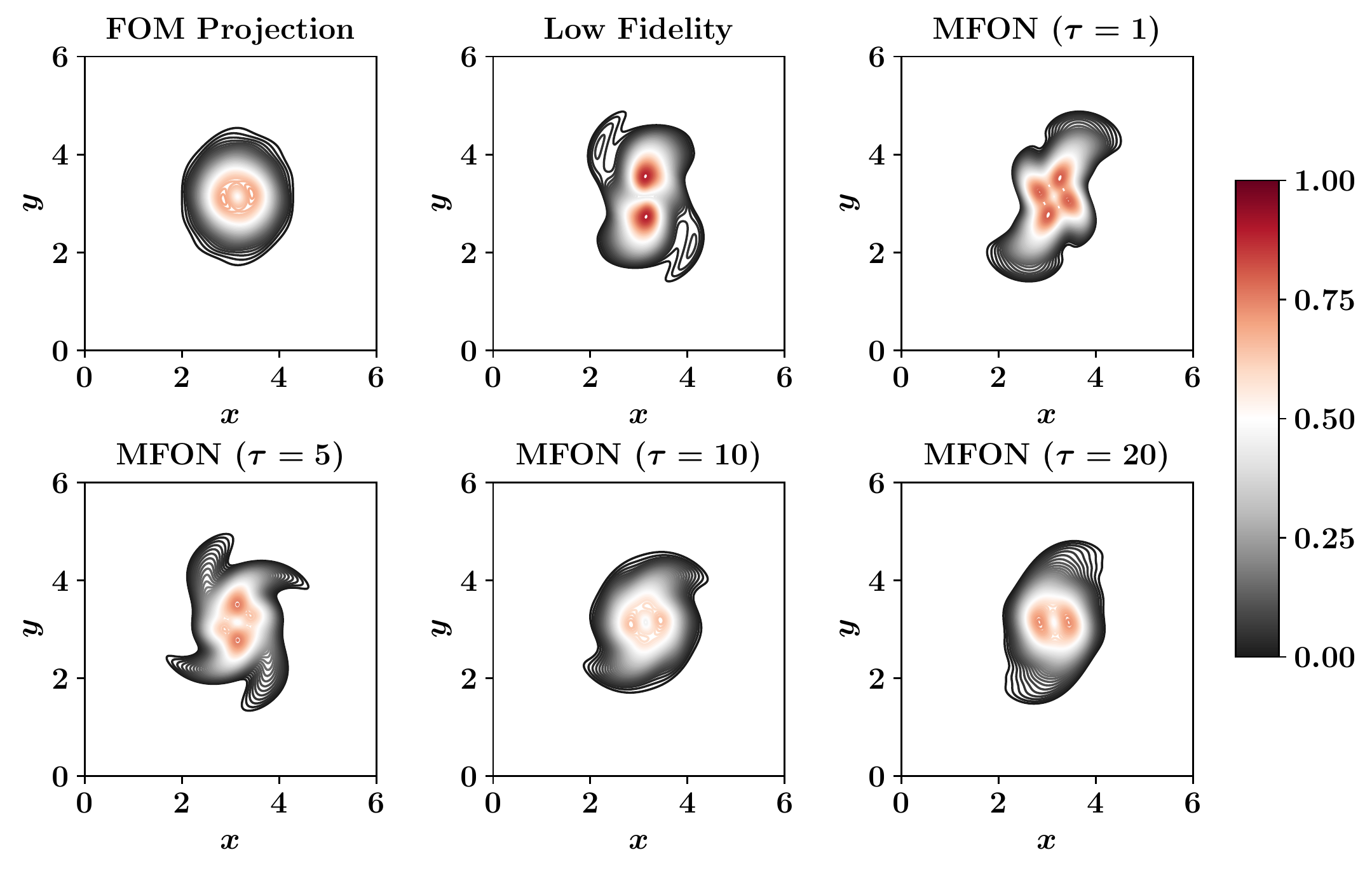}
    \caption{The predicted velocity field at $t=40$ for the vortex merger problem with an initial orientation with $\theta=45^{\circ}$ at $\text{Re} = 3000$, {corresponding to an extrapolative test case}. For in-the-loop training, true data points are assumed to be available also at the intermediate steps within the $\tau$ time steps.}
    \label{fig:vortex_vort_extrapolation_more}
    \end{figure}

The contour plots of reconstructed vorticity fields are presented in \cref{fig:vortex_vort_extrapolation_more}, where we can visually inspect that MFON predictions with $\tau=10$ and $\tau=20$ are the closest to the projection of the FOM solution. Thus, viewing the closure problem as a multifidelity learning problem gives promising results. Moreover, using DeepONet framework allows the extrapolation to new parameter/time regimes where traditional PROMs usually fail. It is worth noting here that the POD bases are constructed with data up to $t=20$, which are less representative of solution at $t=40$ due to the convective nature of the vortex merger problem. Hence, we can spot a relatively large discrepancy between the FOM solution (e.g., \cref{fig:wfom}) and the FOM projection in \cref{fig:vortex_vort_extrapolation_more}. Although the representability of the ROM basis functions is beyond the scope of the current work, we refer the interested readers to recent works on adaptive PROMs by online sampling from FOM \cite{peherstorfer2020model,huang2023predictive}, time-dependent subspaces \cite{patil2020real,ramezanian2021fly}, and custom-made basis function using DeepONets \cite{meuris2023machine}.

\section{Concluding remarks and future work} \label{sec:conclusion}
We have introduced a multifidelity operator learning framework for closure modeling in multiscale systems. We have used the combination of proper orthogonal decomposition (POD) and Galerkin projection to define the low fidelity model and trained a deep neural network (DeepONet) to learn correction terms. We have demonstrated that augmenting the Galerkin POD (GPOD) models with multifidelity operator learning (MFON) improves its generalizability to make predictions for situations with  varying parameters and initial conditions. The extrapolation to different initial conditions is particularly a key advantage compared to state-of-the-art projection-based reduced order models (PROMs). Furthermore, we have leveraged differentiable programming tools to enable in-the-loop training of MFONs and provide a feedback loop between the MFON predictions at one time step to the inputs at the next step. In-the-loop training can be seen as a way of imposing temporal causality in time-dependent problems. Two test cases of convection-dominated flow problems have been considered which corroborate the efficacy of MFONs with in-the-loop training for closure modeling. Our numerical results support the conclusion that exposing the MFON to its own output during the training phase leads to more accurate predictions and expands the predictive skill horizon.

As promising as the results are, many research questions remain open. As detailed in \cref{sec:result}, it is still needed to understand the effect of the time window's length for in-the-loop training and how to optimize it with regard to the specific problem in hand, the size of the training data, the complexity of architecture, etc. {Analogies between the time window for in-the-loop training and the assimilation window in the four-dimensional variational (4D-VAR) data assimilation can potentially provide insights into the selection of $\tau$.} It is also important to analyze the interplay between the low fidelity model of resolved scales and the MFON. Questions like: (1) how do inaccuracies and uncertainties in the low fidelity model feed back into the MFON predictions of the closure and vice versa; (2) how does the error at one time step propagate to the next steps and how does it affect the predictability limits of MFONs; and (3) how does improving/worsening the low fidelity model (e.g., by including more or fewer modes) change the MFON predictions, should be answered. Although we demonstrate the use of MFON to account for the effect of truncated scales on the dynamics of resolved ones in PROM setting, the closure problem manifests itself in an array of applications that can take advantage of the MFON framework. This could be due to coarsening the model resolution to the scales of interest (as in large eddy simulations), parameterization of physical processes (e.g., clouds) in climate modeling, or using physical assumptions to derive simplified governing equations and/or analytical solutions (e.g., self-similarity solutions of PDEs).

\section*{Acknowledgments}
We would like to thank Dr Amanda Howard for helpful discussions and comments. The work of SA is supported by the Department of Energy (DOE) Office of Advanced Scientific Computing Research (ASCR) through the Pacific Northwest National Laboratory Distinguished  Fellowship in Scientific Computing (Project No. 71268). The work of PS was supported by DOE’s Scientific Discovery through Advanced Computing (SciDAC) program via a partnership on Earth system model development between the Office of Biological and Environmental Research (BER) and ASCR (Project No. 79699). Pacific Northwest National Laboratory is operated by Battelle Memorial Institute for DOE under Contract DE-AC05-76RL01830.


\bibliographystyle{unsrtnat}
\bibliography{references}  

\end{document}